\documentclass[11pt]{article}
\usepackage{epsfig} 
\usepackage{amssymb}
\usepackage{graphics}
\newcommand{\sect}[1]{ \section{#1} \setcounter{equation}{0} } 

\newcommand{\half}{\mbox{\small{$\frac{1}{2}$}}}

\newcommand{\Nc}{N_{\!c}}
\newcommand{\Nf}{N_{\!f}}

\newcommand{\bare}{\mbox{\footnotesize{o}}}

\newcommand{\pslash}{p \! \! \! /}
\newcommand{\Dslash}{D \! \! \! \! /}
\newcommand{\MSbar}{\overline{\mbox{MS}}}

\newcommand{\RI}{\mbox{RI${}^\prime$}}

\begin{document}

\title{Four loop Green's functions involving the $n$~$=$~$2$ moment of the 
Wilson operator}

\author{J.A. Gracey, \\ Theoretical Physics Division, \\ 
Department of Mathematical Sciences, \\ University of Liverpool, \\ P.O. Box 
147, \\ Liverpool, \\ L69 3BX, \\ United Kingdom.} 

\date{}

\maketitle 

\vspace{5cm} 
\noindent 
{\bf Abstract.} We evaluate the Green's function for the insertion of the
second moment of the twist-$2$ flavour nonsinglet Wilson operator in a quark 
$2$-point function in all three different single scale external momentum 
configurations at four loops in the $\MSbar$ scheme and the chiral limit. One 
configuration is where the operator is inserted at zero momentum while the 
other two are where a non-zero momentum flows out through the operator itself
with one external quark momentum nullified. In the latter two configurations
mixing of the operator with a total derivative twist-$2$ operator is included
for renormalization group consistency. In addition we compute the correlation 
functions of both gauge invariant operators to four loops in the same scheme.

\vspace{-17.2cm}
\hspace{13.4cm}
{\bf LTH 1387}

\newpage 

\sect{Introduction.}

The dynamics of the quarks internal to the proton and other hadrons are 
governed by parton distribution functions and generalized parton distribution
functions. These have been subject to intense study over many years since, for
instance, they are important in determining cross-sections in hadron physics.
In terms of the quantum field theory construct that underlies the parton 
dynamics, the distribution functions can be accessed from non-forward operator 
matrix elements involving in the first instance the key twist-$2$ gauge 
invariant Wilson operators that describe the dominant scattering process where 
the core theory is Quantum Chromodynamics (QCD). For example for experiments 
the leading twist operators are classified as being either flavour non-singlet 
or singlet as well as whether the scattering is unpolarized or polarized. The 
latter property allows access to the spin dynamics of the hadron. Theoretically
the evolution or renormalization group running of the operators is deduced 
through perturbative calculations. For instance the flavour non-singlet, 
twist-$2$ unpolarized anomalous dimensions are known as a function of the 
operator moment $n$, together with their Mellin transformed splitting function 
partners, to three loops explicitly in the modified minimal subtraction 
($\MSbar$) scheme, \cite{1,2,3}, with these results being verified in an
off-shell context in \cite{4}. Subsequently the four loop structure of the same
quantities has been examined from several directions. For instance four loop 
$\MSbar$ low moment operator dimensions have been recorded in \cite{5,6,7,8}, 
in the large $\Nf$ limit at the same order, \cite{9}, or beyond, \cite{10}, 
where $\Nf$ is the number of quarks. Impressively there has even been progress 
for low moments at five loops, \cite{11}. The three loop expressions of 
\cite{3,4} represent the extent to which full analytic results are available. 
However in recent years significant progress has been made to extend our 
knowledge of the four loop operator anomalous dimensions. Indeed these advances
in multiloop computations in the last decade or so have been made as a
consequence of improved algorithms for four loop Feynman integral evaluation 
and especically via the {\sc Forcer} package, \cite{12,13}, written in the 
symbolic language {\sc Form}, \cite{14,15}. The other ingredient for 
understanding distribution functions is the operator matrix element evaluated 
across all momentum scales. This is a highly non-perturbative quantity that can
only be determined via lattice field theory methods. This description applies 
equally to generalized parton distribution functions where the distinction in 
the type of matrix element needed is simple to define. Non-forward matrix 
elements are Green's functions where the relevant Wilson operator is inserted 
at non-zero momentum in a quark $2$-point function where non-zero momentum 
flows through the two external fields. For the forward case the momentum 
configuration differs in that there is no outwards momentum flow at the 
operator insertion itself. In the former case there are two independent 
external momenta while in the latter there is only one momentum scale.

Although the various matrix elements are well-defined and computed 
non-perturbatively using lattice field theory they ultimately have to match 
onto the high energy or perturbative regime of the same but continuum spacetime
evaluated matrix elements in the massless or chiral limit. The interplay 
between lattice and precision methods in this area can be seen for example in
the non-exhaustive representative set of articles \cite{16,17,18,19,20,21}.
Clearly the more accurately or precisely determined the matrix element is 
perturbatively then the more reliable will be the extrapolation to the lattice 
evaluation. Moreover high order perturbative data will play an important role 
in understanding and quantifying uncertainties. In the last twenty or so years 
there has been a sizeable number of perturbative computations of matrix 
elements. For instance for insight into initial activity in continuum and
lattice perturbation theory related to the forward situation for a few low 
moments see \cite{22}. The continuum perturbative results were extended to 
three loops in \cite{23,24} for moments $n$~$=$~$2$ and $3$ of the forward
matrix element for the unpolarized flavour non-singlet situation. One issue 
that is related to matrix element matching for either set of distribution 
functions is that of renormalization scheme. Perturbatively the canonical 
scheme used is that of $\MSbar$ primarily as one can evaluate massless Feynman 
diagrams contributing to the perturbative matrix element for a single momentum 
flow to the highest loop order possible at present. Indeed for the three loop 
work the {\sc Mincer} package, \cite{25}, was the workhorse automatic Feynman 
graph computation algorithm. Only in recent years has the evaluation of 
non-forward matrix elements for Wilson operators reached the same level of 
three loop accuracy, \cite{26,27,28}. Latterly however {\sc Mincer} has been 
superseded by an algorithm that carries out the same task of evaluating 
massless $2$-point functions in $d$~$=$~$4$~$-$~$2\epsilon$ dimensions but now 
at {\em four} loops. The package known as {\sc Forcer}, \cite{12,13}, is also 
written in the powerful and efficient symbolic manipulation language 
{\sc Form}, \cite{14,15}. Given this advance for perturbative computations it 
is therefore the purpose of this article to evaluate the operator matrix 
element for the twist-$2$ flavour non-singlet Wilson operator with moment 
$n$~$=$~$2$ in the forward configuration extending the three loop work of 
\cite{23}. In addition to provide a not unrelated connection with the 
non-forward setup we will evaluate the matrix element where the single external
momentum flows through the operator itself and one of the two quark legs. We 
will carry this out for both possible cases. Clearly the matrix element will be
gauge variant and hence depend on the parameter $\alpha$ of the linear 
covariant gauge which we use throughout. Although lattice evaluations are 
performed in the Landau gauge, we will retain a non-zero $\alpha$ partly as it 
acts as a tool for consistency checks. It is worth noting that lattice 
measurements for gauge invariant operators are invariably carried out in 
schemes other than $\MSbar$. One such scheme is the regularization invariant 
($\RI$) scheme which was introduced in \cite{29,30}. Lattice measurements in 
this scheme can then be converted to the $\MSbar$ scheme as the relations 
between the field anomalous dimensions and coupling constant in both schemes 
are known to high loop order, \cite{31,32,33}. To assist with the lattice 
matching process we will decompose the matrix elements into a complete basis of
Lorentz tensors and compute each form factor. This will contribute to the 
lattice measurements of the same quantity in the $\RI$ scheme by covering all 
external momentum choices to disentangle the form factor.

While such perturbative results for matrix elements will prove useful for 
matching to the lattice evaluation of the same quantity, we will also compute a
gauge independent Green's function involving the gauge invariant operator. This
is the $2$-point correlation function of the operator which will be evaluated 
to four loops in the $\MSbar$ scheme extending the three loop analysis of 
\cite{34}. The advantage of considering such a correlation function is that it 
can in principle be measured on the lattice without the need to use a gauge 
fixed Lagrangian. In outlining the Green's functions of interest to this study 
we recall that the twist-$2$ $n$~$=$~$2$ moment operator mixes with another 
operator of the same dimension and quantum numbers. Although this is a total 
derivative operator it will not affect the computation of the forward matrix 
element where there is no external momentum exiting the operator insertion. 
However mixing does become crucial where a momentum flows through the operator 
itself. This is particularly important for the other Green's functions 
mentioned earlier as well since mixing cannot be neglected. If it was then 
inconsistencies would arise in the renormalization of the Green's functions as 
the associated renormalization constants could not be correctly converted to 
operator anomalous dimensions. Consequently providing data on the non-forward
matrix element ought to refine the uncertainty in parton distribution 
properties.

The article is organized as follows. In section 2 we recall the core elements
of the formalism surrounding the renormalization of the $n$~$=$~$2$ moment
gauge invariant Wilson operators. Subsequently the determination of the four
loop operator matrix elements in different single momentum flow configurations
are provided in Section 3 with the details of the gauge invariant operator 
correlation computation recorded in Section 4. Concluding remarks are given in 
Section 5. Finally Appendix A contains explicit four loop expressions for the 
asymmetric momentum routing when operator mixing is present.

\sect{Formalism.}

As our study will concern the evaluation of Green's functions of two twist-$2$ 
flavour non-singlet gauge invariant Wilson operators we need to recall the 
basic renormalization properties. First the two independent rank two operators 
we take are
\begin{equation}
{\cal O}_1^{\mu\nu} ~ \equiv ~ 
{\cal S} \bar{\psi} \gamma^\mu D^\nu \psi ~~~,~~~
{\cal O}_2^{\mu\nu} ~ \equiv ~ {\cal S} \partial^\mu \left( \bar{\psi} 
\gamma^{\nu} \psi \right) 
\end{equation}
where the open flavour indices on the quarks are omitted and the symmetrization
and tracelessness operator ${\cal S}$ acts on the Lorentz indices meaning the 
explicit combination of each core operator is
\begin{eqnarray}
{\cal O}_1^{\mu\nu} &=& 
\bar{\psi} \gamma^\mu D^\nu \psi ~+~ \bar{\psi} \gamma^\nu D^\mu \psi ~-~
\frac{2}{d} \eta^{\mu\nu} \bar{\psi} \Dslash \psi \nonumber \\
{\cal O}_2^{\mu\nu} &=& 
\partial^\mu \left( \bar{\psi} \gamma^{\nu} \psi \right) ~+~
\partial^\nu \left( \bar{\psi} \gamma^{\mu} \psi \right) ~-~ \frac{2}{d} 
\eta^{\mu\nu} \partial_\sigma \left( \bar{\psi} \gamma^{\sigma} \psi \right) ~.
\end{eqnarray}
A third operator is possible which is
${\cal S} \left( D^\mu \bar{\psi} \right) \gamma^\nu \psi$ but it is 
straightforward to show that this is a linear combination of 
${\cal O}_1^{\mu\nu}$ and ${\cal O}_2^{\mu\nu}$ using integration by parts.
The total derivative operator is based on the lowest rank twist-$2$ operator
$\bar{\psi} \gamma^\mu \psi$ and is physical meaning it has zero anomalous
dimension. We have made this particular operator basis choice to be consistent 
with previous work \cite{23}. For Green's functions involving 
${\cal O}_1^{\mu\nu}$ where there is no external momentum flowing through this
operator one only requires its associated multiplicative renormalization 
constant. However in general the operator mixes with ${\cal O}_2^{\mu\nu}$
under renormalization, even in the chiral limit, since they both have the same 
dimension and quantum numbers. The symmetrization and tracelessness properties 
ensure there is no mixing with other twist-$2$ operators provided the quarks 
are massless. The mixing of both operators becomes significant when either are 
present in a Green's function and an external momentum flows through the 
operator insertion. As this will arise in several of the Green's functions we 
will evaluate we recall the relevant formalism.

First the relation between the bare and renormalized operators is given by
\begin{equation}
{\cal O}_{i\,{\bare}} ~=~ Z_{ij} {\cal O}_j
\end{equation}
where ${}_{\bare}$ will denote a bare entity throughout. For our basis choice
the matrix of renormalization constants, $Z_{ij}$, will be upper triangular
taking the form
\begin{equation}
Z_{ij} ~=~ \left(
\begin{array}{cc}
Z_{11} & Z_{12} \\
0 & Z_{22} \\
\end{array}
\right) ~.
\label{opmixmat}
\end{equation}
For example $Z_{11}$ is the multiplicative renormalization constant for
${\cal O}_1^{\mu\nu}$ itself when it is inserted at zero momentum in a Green's
function. The matrix of operator anomalous dimensions are defined in a standard
way via
\begin{equation}
\gamma_{ij}(a) ~=~ \mu \frac{d ~}{d \mu} \ln Z_{ij}
\label{mixmatdef}
\end{equation}
where
\begin{equation}
\mu \frac{d~}{d\mu} ~=~ \beta(a) \frac{\partial ~}{\partial a} ~+~
\alpha \gamma_\alpha(a,\alpha) \frac{\partial ~}{\partial \alpha} 
\label{muderive}
\end{equation}
and $a$ is related to the gauge coupling constant $g$ by 
$a$~$=$~$g^2/(16\pi^2)$. We have included the anomalous dimension of the gauge 
parameter $\alpha$ of the canonical linear covariant gauge in this differential
operator in order to be complete. However throughout we will renormalize the 
gauge independent operators in the $\MSbar$ scheme meaning that the operator 
anomalous dimensions will be $\alpha$ independent, \cite{35}. So the argument 
of the operator anomalous dimensions and the $\beta$-function will only involve
$a$. By contrast a set of the Green's functions considered here are gauge 
variant meaning they themselves will be $\alpha$ dependent. We retain 
$\alpha$~$\neq$~$0$ throughout partly to be as general as possible but also
because the parameter plays a useful role in the internal checking procedures 
for high order computations. Ultimately though the main usefulness of the 
Green's functions will be for the Landau gauge lattice matching which will 
follow as a corollary by setting $\alpha$~$=$~$0$. More concretely once the 
elements of (\ref{mixmatdef}) are available to a particular loop order 
explicitly then the operator anomalous dimensions are deduced from
\begin{eqnarray}
0 &=& \gamma_{11}(a) Z_{11} ~+~ \mu \frac{d ~}{d \mu} Z_{11} \nonumber \\
0 &=& \gamma_{11}(a) Z_{12} ~+~ \gamma_{12}(a) Z_{22} ~+~
\mu \frac{d ~}{d \mu} Z_{12} \nonumber \\
0 &=& \gamma_{22}(a) Z_{22} ~+~ \mu \frac{d ~}{d \mu} Z_{22} 
\end{eqnarray}
in general.

In practical terms the three renormalization constants can be determined by
considering a set of $3$-point functions where there is only one external
momentum as opposed to the maximum number of independent momenta which is two. 
For ${\cal O}_i^{\mu\nu}$ since the lowest number of fields in each is two the 
Green's function is that where the operator is inserted in a quark $2$-point 
function, \cite{1,2,3,4}. More explicitly the three possibilities are
\begin{eqnarray}
G^{\mu\nu}_i (p) &=& \langle \psi(p) ~ 
{\cal O}_i^{\mu\nu}(0) ~ \bar{\psi} (-p) \rangle \\
\widehat{G}^{\mu\nu}_i (p) &=& \langle \psi(p) ~ 
{\cal O}_i^{\mu\nu}(-p) ~ \bar{\psi} (0) \rangle \\
\widetilde{G}^{\mu\nu}_i (p) &=& \langle \psi(0) ~ 
{\cal O}_i^{\mu\nu}(-p) ~ \bar{\psi} (p) \rangle 
\label{opinsgf}
\end{eqnarray}
where the computation of $G^{\mu\nu}_i (p)$ will determine $Z_{11}$ and
$Z_{22}$. We will refer to this external momentum configuration as the 
symmetric one. The final renormalization constant $Z_{12}$ is extracted from 
evaluating either of the two remaining Green's functions once the other two
are available to the same loop order. These will be regarded as the asymmetric
configurations. For the purposes of assisting lattice analyses we will 
determine all three Green's functions to four loops in the $\MSbar$ scheme.

\sect{Operator matrix elements.}

Having identified the general properties of the two operators, in this section 
we turn to the more practical task of determining the Green's functions of 
(\ref{opinsgf}) to four loops in the $\MSbar$ scheme. The first case we 
consider is where the operators are inserted in a quark $2$-point function with
a single momentum flow. The key computational technique is to apply the 
{\sc Forcer} algorithm, \cite{12,13}, that evaluates massless $2$-point 
functions to four loops in dimensional regularization in 
$d$~$=$~$4$~$-$~$2\epsilon$ dimensions. The {\sc Forcer} package is written in 
the symbolic manipulation language {\sc Form} \cite{14,15} which is the core 
workhorse that effected all the main computations in this article. The 
electronic representation of the Feynman graphs contributing to each Green's 
function were generated by {\sc Qgraf}, \cite{36}, with the number of graphs 
for each loop order and Green's function given in Table \ref{opinsnum}. As 
{\sc Forcer} requires Lorentz scalar Feynman graphs the initial step is to 
decompose (\ref{opinsgf}) into a complete tensor basis and to fix our notation 
we set
\begin{eqnarray}
G^{\mu\nu}_i (p) &=& \sum_{j=1}^2 \Sigma^{(j)}_{i}(p) {\cal T}_j^{\mu\nu}(p) 
\nonumber \\
\widehat{G}^{\mu\nu}_i (p) &=& 
\sum_{j=1}^2 \widehat{\Sigma}^{(j)}_{i}(p) {\cal T}_j^{\mu\nu}(p) \nonumber \\
\widetilde{G}^{\mu\nu}_i (p) &=& 
\sum_{j=1}^2 \widetilde{\Sigma}^{(j)}_{i}(p) {\cal T}_j^{\mu\nu}(p)
\end{eqnarray}
for $i$~$=$~$1$ and $2$ where the two basis tensors are, \cite{23},
\begin{equation}
{\cal T}_1^{\mu\nu}(p) ~=~ 
\gamma^\mu p^\nu ~+~ \gamma^\nu p^\mu ~-~ \frac{2}{d} \pslash \eta^{\mu\nu}
~~~,~~~ 
{\cal T}_2^{\mu\nu}(p) ~=~ \left[ p^\mu p^\nu ~-~ \frac{p^2}{d} \eta^{\mu\nu} 
\right] \pslash 
\end{equation}
which are symmetric and traceless consistent with the same property of
${\cal O}_i^{\mu\nu}$. This basis choice is complete and sufficient for the 
lattice to construct a specific $\RI$ scheme for the operator matrix element 
measurement itself. By construction it is trivial to see that 
$G^{\mu\nu}_2(p)$~$=$~$0$ to all orders. Equally 
$\widehat{G}^{\mu\nu}_2(p)$~$=$~$\widetilde{G}^{\mu\nu}_2(p)$ by symmetry 
meaning that only four Green's functions need to be determined. The respective 
form factors can be deduced from the projections, \cite{23},
\begin{eqnarray} 
\Sigma^{(1)}_i(p) &=& \frac{1}{8(d-1)} \left[ \frac{}{}
\mbox{tr} \left( {\cal T}_1^{\mu\nu}(p) G_{i\,\mu\nu}(p) \right) ~-~ 
2 \, \mbox{tr} \left( {\cal T}_2^{\mu\nu}(p) G_{i\,\mu\nu}(p) \right) \right] 
\nonumber \\  
\Sigma^{(2)}_i(p) &=& -~ \frac{1}{4(d-1)} \left[ \frac{}{} 
\mbox{tr} \left( {\cal T}_1^{\mu\nu}(p) G_{i\,\mu\nu}(p) \right) ~-~ 
(d+2) \, \mbox{tr} \left( {\cal T}_2^{\mu\nu}(p) G_{i\,\mu\nu}(p) \right) 
\right]  
\label{omedef}
\end{eqnarray}
with similar decompositions for the other two momenta routings via the formal 
respective replacements
$\{ G^{(k)}_{i\,\mu\nu}(p),\Sigma^{(k)}_i(p)\}$~$\rightarrow$~$\{ \widehat{G}^{(k)}_{i\,\mu\nu}(p), \widehat{\Sigma}^{(k)}_i(p)\}$ and
$\{ G^{(k)}_{i\,\mu\nu}(p),\Sigma^{(k)}_i(p)\}$~$\rightarrow$~$\{ \widetilde{G}^{(k)}_{i\,\mu\nu}(p), \widetilde{\Sigma}^{(k)}_i(p)\}$
where $\mbox{tr}$ is the spinor trace.

{\begin{table}[ht]
\begin{center}
\begin{tabular}{|c||r|r|r|}
\hline
\rule{0pt}{12pt}
$L$ & $G^{\mu\nu}_1(p)$ & $\widehat{G}^{\mu\nu}_1(p)$ &
$\widehat{G}^{\mu\nu}_2 (p)$ \\
\hline
$1$ & $3$ & $3$ & $1$ \\
$2$ & $32$ & $32$ & $13$ \\
$3$ & $566$ & $566$ & $244$ \\
$4$ & $12878$ & $12878$ & $5728$ \\
\hline
Total & $13479$ & $13479$ & $5986$ \\
\hline
\end{tabular}
\end{center}
\begin{center}
\caption{Number of graphs at each loop order $L$ for operator insertions in
quark $2$-point functions for two different external momenta routings.}
\label{opinsnum}
\end{center}
\end{table}}

Before recording our results the first stage is the verification of the four
loop operator flavour nonsinglet anomalous dimensions which were computed 
originally in \cite{1,2,3,4,5,6}. The anomalous dimension of 
${\cal O}_1^{\mu\nu}$ is
\begin{eqnarray}
\gamma_{11}(a) &=&
\frac{8}{3} C_F a
+ \left[
\frac{376}{27} C_A C_F
- \frac{128}{27} C_F T_F \Nf
- \frac{112}{27} C_F^2
\right] a^2 \nonumber \\
&&
+~ \left[
\frac{64}{3} \zeta_3 C_A^2 C_F
+ \frac{128}{3} \zeta_3 C_F^2 T_F \Nf
+ \frac{128}{3} \zeta_3 C_F^3
+ \frac{20920}{243} C_A^2 C_F
- \frac{8528}{243} C_A C_F^2
\right. \nonumber \\
&& \left. ~~~~~
-~ \frac{6824}{243} C_F^2 T_F \Nf
- \frac{6256}{243} C_A C_F T_F \Nf
- \frac{896}{243} C_F T_F^2 \Nf^2
- \frac{560}{243} C_F^3
\right. \nonumber \\
&& \left. ~~~~~
-~ \frac{128}{3} \zeta_3 C_A C_F T_F \Nf
- 64 \zeta_3 C_A C_F^2
\right] a^3 \nonumber \\
&&
+~ \left[
352 \zeta_4 C_A^2 C_F^2
- \frac{1626064}{2187} C_A^2 C_F^2
- \frac{355496}{2187} C_A C_F^2 T_F \Nf
- \frac{106036}{243} C_A^2 C_F T_F \Nf
\right. \nonumber \\
&& \left. ~~~~~
-~ \frac{25744}{27} \zeta_3 C_A^2 C_F^2
- \frac{12160}{27} \zeta_5 C_A^3 C_F
- \frac{8192}{2187} C_F T_F^3 \Nf^3
- \frac{8080}{9} \zeta_3 C_A^2 C_F T_F \Nf
\right. \nonumber \\
&& \left. ~~~~~
-~ \frac{5056}{81} \zeta_3 C_F^3 T_F \Nf
- \frac{2560}{9} \zeta_5 \Nf \frac{d_F^{abcd} d_F^{abcd}}{\Nc}
- \frac{1280}{3} \zeta_5 C_F^3 T_F \Nf
- \frac{1280}{3} \zeta_5 C_F^4
\right. \nonumber \\
&& \left. ~~~~~
-~ \frac{1088}{3} \zeta_4 C_A C_F^2 T_F \Nf
- \frac{736}{9} \frac{d_F^{abcd} d_A^{abcd}}{\Nc}
- \frac{704}{3} \zeta_4 C_A C_F^3
- \frac{512}{3} \zeta_3 C_F^2 T_F^2 \Nf^2
\right. \nonumber \\
&& \left. ~~~~~
-~ \frac{352}{3} \zeta_4 C_A^3 C_F
- \frac{256}{3} \zeta_4 C_A C_F T_F^2 \Nf^2
+ \frac{256}{3} \zeta_4 C_F^2 T_F^2 \Nf^2
+ \frac{256}{3} \zeta_4 C_F^3 T_F \Nf
\right. \nonumber \\
&& \left. ~~~~~
+~ \frac{512}{3} \zeta_3 C_A C_F T_F^2 \Nf^2
+ \frac{640}{9} \zeta_5 C_A C_F^2 T_F \Nf
+ \frac{832}{3} \zeta_4 C_A^2 C_F T_F \Nf
\right. \nonumber \\
&& \left. ~~~~~
+~ \frac{832}{9} \Nf \frac{d_F^{abcd} d_F^{abcd}}{\Nc}
+ \frac{1024}{9} \zeta_3 \Nf \frac{d_F^{abcd} d_F^{abcd}}{\Nc}
+ \frac{1024}{81} \zeta_3 C_F T_F^3 \Nf^3
\right. \nonumber \\
&& \left. ~~~~~
+~ \frac{1280}{3} \zeta_5 C_A C_F^3
+ \frac{1984}{9} \zeta_3 \frac{d_F^{abcd} d_A^{abcd}}{\Nc}
+ \frac{4480}{9} \zeta_5 C_A^2 C_F^2
+ \frac{5120}{9} \zeta_5 \frac{d_F^{abcd} d_A^{abcd}}{\Nc}
\right. \nonumber \\
&& \left. ~~~~~
+~ \frac{8960}{27} \zeta_5 C_A^2 C_F T_F \Nf
+ \frac{10880}{81} \zeta_3 C_F^4
+ \frac{25400}{729} C_A C_F T_F^2 \Nf^2
+ \frac{25856}{27} \zeta_3 C_A C_F^2 T_F \Nf
\right. \nonumber \\
&& \left. ~~~~~
+~ \frac{31040}{81} \zeta_3 C_A C_F^3
+ \frac{34936}{81} \zeta_3 C_A^3 C_F
+ \frac{99776}{2187} C_F^2 T_F^2 \Nf^2
+ \frac{194392}{2187} C_F^4
\right. \nonumber \\
&& \left. ~~~~~
+~ \frac{238676}{2187} C_A C_F^3
+ \frac{381824}{2187} C_F^3 T_F \Nf
+ \frac{1734130}{2187} C_A^3 C_F
\right] a^4 ~+~ O(a^5)
\label{gamma11}
\end{eqnarray}
to four loops and is provided to allow orientation of our notation to that of
others such as \cite{5,6}. In (\ref{gamma11}) $\zeta_n$ is the Riemann zeta 
function, $C_F$, $C_A$ and $T_F$ are the usual rank $2$ Casimirs, $\Nc$ is the
dimension of the fundamental representation and
\begin{equation}
d_R^{abcd} ~=~ \frac{1}{6} \mbox{Tr} \left( T^a T^{(b} T^c T^{d)} \right)
\end{equation}
is the rank four totally symmetric group trace of the group generators in 
representation $R$ with $A$ and $F$ denoting the adjoint and fundamental
representations respectively, \cite{37}. We note that (\ref{gamma11}) agrees 
with the first four loop non-singlet operator dimension calculated in
\cite{5,6}. In \cite{5} the $SU(3)$ expression for $\Nf$~$=$~$3$ was recorded 
and restricting (\ref{gamma11}) to the same value reproduces equations (9) to 
(11) of that article. The full explicit expression acts as a check on the 
finite part of the Green's functions which is our main goal. In addition we 
have verified by explicit computation that
\begin{equation}
\gamma_{12}(a) ~=~ -~ \half \gamma_{11}(a) ~+~ O(a^5)
\label{gamma12}
\end{equation}
is satisfied to four loops. This can be deduced generally by recalling that the
operator ${\cal S} \left( D^\mu \bar{\psi} \right) \gamma^\nu \psi$ is related
to ${\cal O}_1$ and ${\cal O}_2$ by integration by parts. Since 
${\cal S} \left( D^\mu \bar{\psi} \right) \gamma^\nu \psi$ will have the same
anomalous dimension then (\ref{gamma12}) follows. The relation has arisen in
lattice field theory \cite{22} and discussed in the generalization of the
relation to higher moments, \cite{38}. Finally we have also explicitly verified
that
\begin{equation}
\gamma_{22}(a) ~=~ O(a^5)
\label{gamma22}
\end{equation}
reflecting the fact that the rank one twist-$2$ operator itself is physical
with zero anomalous dimension meaning ${\cal O}_2^{\mu\nu}$ inherits the same
property.

Having reproduced the four loop $\MSbar$ expression for $\gamma_{11}(a)$ in 
full agreement with \cite{1,2,3,4,5,6} we can now record the expressions for 
the form factors to the same order. For the zero momentum insertion of
${\cal O}_1^{\mu\nu}$ we have
\begin{eqnarray} 
\left. \Sigma^{(1)}_1 (p) \right|_{p^2 = \mu^2} ^{\alpha = 0} &=& 1
- \frac{31}{9} C_F a \nonumber \\
&&
+ \left[
8 \zeta_3 C_F C_A
- 8 \zeta_3 C_F^2
- \frac{9487}{324} C_F C_A
+ \frac{2101}{162} \Nf T_F C_F
+ \frac{8195}{648} C_F^2
\right] a^2 \nonumber \\
&&
+ \left[
\frac{135}{4} \zeta_5 C_F C_A^2
+ \frac{400}{3} \zeta_5 C_F^3
- \frac{21026833}{69984} C_F C_A^2
- \frac{63602}{2187} \Nf^2 T_F^2 C_F
\right. \nonumber \\
&& \left. ~~~~
-~ \frac{7964}{81} \zeta_3 C_F^3
- \frac{719}{48} \zeta_4 C_F C_A^2
- \frac{440}{3} \zeta_5 C_F^2 C_A
- \frac{272}{9} \zeta_3 \Nf T_F C_F^2
\right. \nonumber \\
&& \left. ~~~~
-~ \frac{256}{81} \zeta_3 \Nf^2 T_F^2 C_F
- \frac{64}{3} \zeta_4 C_F^3
- \frac{64}{3} \zeta_4 \Nf T_F C_F^2
+ \frac{64}{3} \zeta_4 \Nf T_F C_F C_A
\right. \nonumber \\
&& \left. ~~~~
+~ \frac{439}{9} \zeta_3 C_F^2 C_A
+ \frac{976}{81} \zeta_3 \Nf T_F C_F C_A
+ \frac{2935}{5832} C_F^3
+ \frac{3230}{81} \zeta_3 C_F C_A^2
\right. \nonumber \\
&& \left. ~~~~
+~ \frac{61322}{2187} \Nf T_F C_F^2
+ \frac{452579}{2187} \Nf T_F C_F C_A
+ \frac{605431}{4374} C_F^2 C_A
+ 38 \zeta_4 C_F^2 C_A
\right] a^3 \nonumber \\
&&
+ \left[
\frac{500}{9} \zeta_6 \Nf T_F C_F^2 C_A
- \frac{14907640511}{3359232} C_F C_A^3
- \frac{376659359}{314928} \Nf T_F C_F^3
\right. \nonumber \\
&& \left. ~~~~
-~ \frac{326929915}{279936} C_F^4
- \frac{87106757}{157464} \Nf^2 T_F^2 C_F^2
- \frac{12036989}{8748} \Nf^2 T_F^2 C_F C_A
\right. \nonumber \\
&& \left. ~~~~
-~ \frac{8254165}{31104} \zeta_3 C_F C_A^3
- \frac{2738833}{13824} \zeta_4 C_F C_A^3
- \frac{1888636}{729} \zeta_3 \Nf T_F C_F^2 C_A
\right. \nonumber \\
&& \left. ~~~~
-~ \frac{946255}{648} \zeta_5 \Nf T_F C_F C_A^2
- \frac{871619}{1728} \zeta_3^2 C_F C_A^3
- \frac{832735}{972} \zeta_3 C_F^3 C_A
\right. \nonumber \\
&& \left. ~~~~
-~ \frac{243275}{576} \zeta_6 \frac{d_F^{abcd} d_A^{abcd}}{\Nc}
- \frac{174937}{96} \zeta_7 C_F^2 C_A^2
- \frac{108751}{864} \frac{d_F^{abcd} d_A^{abcd}}{\Nc}
\right. \nonumber \\
&& \left. ~~~~
-~ \frac{97913}{27} \zeta_5 C_F^3 C_A
- \frac{84649}{144} \zeta_5 C_F^2 C_A^2
- \frac{64699}{108} \zeta_3 \frac{d_F^{abcd} d_A^{abcd}}{\Nc}
\right. \nonumber \\
&& \left. ~~~~
-~ \frac{57225}{32} \zeta_7 \frac{d_F^{abcd} d_A^{abcd}}{\Nc}
- \frac{39196}{81} \zeta_3 \Nf^2 T_F^2 C_F C_A
- \frac{17234}{3} \zeta_7 C_F^4
\right. \nonumber \\
&& \left. ~~~~
-~ \frac{11680}{27} \zeta_4 \Nf T_F C_F^2 C_A
- \frac{8360}{27} \zeta_5 \Nf \frac{d_F^{abcd} d_F^{abcd}}{\Nc}
\right. \nonumber \\
&& \left. ~~~~
-~ \frac{7040}{9} \zeta_3^2 \Nf \frac{d_F^{abcd} d_F^{abcd}}{\Nc}
- \frac{6950}{27} \zeta_6 \Nf T_F C_F C_A^2
- \frac{5855}{192} \zeta_4 \frac{d_F^{abcd} d_A^{abcd}}{\Nc}
\right. \nonumber \\
&& \left. ~~~~
-~ \frac{3500}{3} \zeta_6 C_F^3 C_A
- \frac{1142}{9} \zeta_4 C_F^4
- \frac{1100}{27} \zeta_3^2 \Nf T_F C_F C_A^2
\right. \nonumber \\
&& \left. ~~~~
-~ \frac{343}{2} \zeta_7 \Nf T_F C_F C_A^2
- \frac{320}{3} \zeta_3^2 \Nf T_F C_F^3
- \frac{128}{3} \zeta_4 \Nf \frac{d_F^{abcd} d_F^{abcd}}{\Nc}
\right. \nonumber \\
&& \left. ~~~~
-~ \frac{128}{27} \zeta_4 \Nf^3 T_F^3 C_F
+ \frac{592}{243} \zeta_3 \Nf^3 T_F^3 C_F
+ \frac{656}{9} \zeta_4 \Nf T_F C_F^3
\right. \nonumber \\
&& \left. ~~~~
+~ \frac{800}{3} \zeta_6 \Nf T_F C_F^3
+ \frac{944}{9} \zeta_3^2 \Nf T_F C_F^2 C_A
+ \frac{1172}{27} \zeta_4 C_F^3 C_A
\right. \nonumber \\
&& \left. ~~~~
+~ \frac{1280}{3} \zeta_3^2 C_F^4
+ \frac{1600}{9} \zeta_6 \Nf \frac{d_F^{abcd} d_F^{abcd}}{\Nc}
+ \frac{1940}{27} \Nf \frac{d_F^{abcd} d_F^{abcd}}{\Nc}
\right. \nonumber \\
&& \left. ~~~~
+~ \frac{2000}{3} \zeta_6 C_F^4
+ \frac{2444}{9} \zeta_5 \Nf T_F C_F^3
+ \frac{5248}{27} \zeta_5 \Nf^2 T_F^2 C_F C_A
\right. \nonumber \\
&& \left. ~~~~
+~ \frac{6130}{9} \zeta_5 \Nf T_F C_F^2 C_A
+ \frac{9373}{32} \zeta_7 C_F C_A^3
+ \frac{12925}{72} \zeta_6 C_F^2 C_A^2
\right. \nonumber \\
&& \left. ~~~~
+~ \frac{16133}{48} \zeta_4 \Nf T_F C_F C_A^2
+ \frac{35663}{108} \zeta_4 C_F^2 C_A^2
+ \frac{39893}{36} \zeta_3^2 C_F^2 C_A^2
\right. \nonumber \\
&& \left. ~~~~
+~ \frac{53456}{27} \zeta_3 \Nf \frac{d_F^{abcd} d_F^{abcd}}{\Nc}
+ \frac{119470}{27} \zeta_5 C_F^4
+ \frac{198983}{243} \zeta_3 \Nf T_F C_F^3
\right. \nonumber \\
&& \left. ~~~~
+~ \frac{363089}{288} \zeta_3^2 \frac{d_F^{abcd} d_A^{abcd}}{\Nc}
+ \frac{392620}{729} \zeta_3 \Nf^2 T_F^2 C_F^2
+ \frac{859739}{6561} \Nf^3 T_F^3 C_F
\right. \nonumber \\
&& \left. ~~~~
+~ \frac{972485}{864} \zeta_5 \frac{d_F^{abcd} d_A^{abcd}}{\Nc}
+ \frac{1085635}{648} \zeta_3 \Nf T_F C_F C_A^2
+ \frac{1178713}{1458} \zeta_3 C_F^4
\right. \nonumber \\
&& \left. ~~~~
+~ \frac{1386025}{6912} \zeta_6 C_F C_A^3
+ \frac{4137037}{11664} \zeta_3 C_F^2 C_A^2
+ \frac{4973957}{2592} \zeta_5 C_F C_A^3
\right. \nonumber \\
&& \left. ~~~~
+~ \frac{36069049}{7776} \Nf T_F C_F C_A^2
+ \frac{339255815}{314928} C_F^2 C_A^2
\right. \nonumber \\
&& \left. ~~~~
+~ \frac{395518639}{314928} \Nf T_F C_F^2 C_A
+ \frac{1445710241}{629856} C_F^3 C_A
\right. \nonumber \\
&& \left. ~~~~
-~ 1176 \zeta_7 \Nf \frac{d_F^{abcd} d_F^{abcd}}{\Nc}
- 892 \zeta_3^2 C_F^3 C_A
- 64 \zeta_5 \Nf^2 T_F^2 C_F^2
\right. \nonumber \\
&& \left. ~~~~
-~ 52 \zeta_4 \Nf^2 T_F^2 C_F C_A
+ 52 \zeta_4 \Nf^2 T_F^2 C_F^2
+ 294 \zeta_7 \Nf T_F C_F^2 C_A
\right. \nonumber \\
&& \left. ~~~~
+~ 4872 \zeta_7 C_F^3 C_A
\right] a^4 ~+~ O(a^5)
\end{eqnarray}
and
\begin{eqnarray}
\left. \Sigma^{(2)}_1 (p) \right|_{p^2 = \mu^2}^{\alpha = 0} &=& 
-~ \frac{2}{3} C_F a \nonumber \\
&&
+ \left[
\frac{20}{3} \Nf T_F C_F
+ \frac{74}{3} C_F^2
- \frac{284}{9} C_F C_A
- 16 \zeta_3 C_F^2
+ 6 \zeta_3 C_F C_A
\right] a^2 \nonumber \\
&&
+ \left[
\frac{1480}{3} \zeta_5 C_F^2 C_A
- \frac{2762093}{3888} C_F C_A^2
- \frac{9680}{243} \Nf^2 T_F^2 C_F
\right. \nonumber \\
&& \left. ~~~~
-~ \frac{7450}{9} \zeta_3 C_F^2 C_A
- \frac{4913}{27} \Nf T_F C_F^2
- \frac{1760}{3} \zeta_5 C_F^3
- \frac{445}{3} \zeta_5 C_F C_A^2
\right. \nonumber \\
&& \left. ~~~~
-~ \frac{184}{3} \zeta_3 \Nf T_F C_F C_A
+ \frac{1088}{9} \zeta_3 \Nf T_F C_F^2
+ \frac{1617}{4} \zeta_3 C_F C_A^2
\right. \nonumber \\
&& \left. ~~~~
+~ \frac{4232}{9} \zeta_3 C_F^3
+ \frac{29483}{972} C_F^3
+ \frac{87980}{243} \Nf T_F C_F C_A
+ \frac{198059}{324} C_F^2 C_A
\right] a^3 \nonumber \\
&&
+ \left[
7056 \zeta_7 \Nf \frac{d_F^{abcd} d_F^{abcd}}{\Nc}
- \frac{10094785265}{559872} C_F C_A^3
\right. \nonumber \\
&& \left. ~~~~
-~ \frac{101483369}{13122} \Nf T_F C_F^2 C_A
- \frac{62388353}{1944} \zeta_3 C_F^2 C_A^2
\right. \nonumber \\
&& \left. ~~~~
-~ \frac{24576347}{13122} \Nf T_F C_F^3
- \frac{7697425}{1728} \zeta_5 C_F C_A^3
\right. \nonumber \\
&& \left. ~~~~
-~ \frac{5583122}{243} \zeta_3 C_F^4
- \frac{2499584}{729} \Nf^2 T_F^2 C_F C_A
- \frac{424501}{64} \zeta_7 C_F C_A^3
\right. \nonumber \\
&& \left. ~~~~
-~ \frac{272237}{54} \zeta_3 \Nf T_F C_F C_A^2
- \frac{132680}{27} \zeta_5 \Nf T_F C_F^2 C_A
\right. \nonumber \\
&& \left. ~~~~
-~ \frac{114688}{243} \zeta_3 \Nf^2 T_F^2 C_F^2
- \frac{38344}{9} \zeta_3 \Nf \frac{d_F^{abcd} d_F^{abcd}}{\Nc}
- \frac{25432}{9} \zeta_3 \Nf T_F C_F^3
\right. \nonumber \\
&& \left. ~~~~
-~ \frac{21413}{16} \zeta_7 \frac{d_F^{abcd} d_A^{abcd}}{\Nc}
- \frac{18976}{3} \zeta_3^2 C_F^4
- \frac{18560}{9} \zeta_5 \Nf \frac{d_F^{abcd} d_F^{abcd}}{\Nc}
\right. \nonumber \\
&& \left. ~~~~
-~ \frac{4887}{2} \zeta_3^2 \frac{d_F^{abcd} d_A^{abcd}}{\Nc}
- \frac{3328}{3} \zeta_3^2 \Nf T_F C_F^3
- \frac{640}{3} \zeta_5 \Nf^2 T_F^2 C_F^2
\right. \nonumber \\
&& \left. ~~~~
-~ \frac{448}{3} \zeta_3^2 \Nf T_F C_F C_A^2
- \frac{128}{9} \zeta_4 \Nf T_F C_F^2 C_A
- \frac{76}{3} \zeta_4 C_F^3 C_A
\right. \nonumber \\
&& \left. ~~~~
+~ \frac{4}{3} \Nf \frac{d_F^{abcd} d_F^{abcd}}{\Nc}
+ \frac{128}{9} \zeta_4 C_F^4
+ \frac{128}{9} \zeta_4 \Nf T_F C_F^3
+ \frac{719}{72} \zeta_4 C_F^2 C_A^2
\right. \nonumber \\
&& \left. ~~~~
+~ \frac{1840}{27} \zeta_5 \Nf^2 T_F^2 C_F C_A
+ \frac{2840}{9} \zeta_3 \Nf^2 T_F^2 C_F C_A
+ \frac{10000}{9} \zeta_5 \frac{d_F^{abcd} d_A^{abcd}}{\Nc}
\right. \nonumber \\
&& \left. ~~~~
+~ \frac{20320}{3} \zeta_5 \Nf T_F C_F^3
+ \frac{38305}{18} \zeta_5 \Nf T_F C_F C_A^2
+ \frac{47408}{3} \zeta_3^2 C_F^3 C_A
\right. \nonumber \\
&& \left. ~~~~
+~ \frac{59011}{24} \zeta_3^2 C_F C_A^3
+ \frac{123871}{72} \zeta_3 \frac{d_F^{abcd} d_A^{abcd}}{\Nc}
+ \frac{194012}{3} \zeta_7 C_F^4
\right. \nonumber \\
&& \left. ~~~~
+~ \frac{201451}{144} \frac{d_F^{abcd} d_A^{abcd}}{\Nc}
+ \frac{303320}{27} \zeta_5 C_F^3 C_A
+ \frac{426775}{216} \zeta_5 C_F^2 C_A^2
\right. \nonumber \\
&& \left. ~~~~
+~ \frac{562720}{2187} \Nf^3 T_F^3 C_F
+ \frac{605871}{16} \zeta_7 C_F^2 C_A^2
+ \frac{1469341}{243} \zeta_3 \Nf T_F C_F^2 C_A
\right. \nonumber \\
&& \left. ~~~~
+~ \frac{3947639}{26244} C_F^3 C_A
+ \frac{4212205}{108} \zeta_3 C_F^3 C_A
+ \frac{4368119}{2916} C_F^4
\right. \nonumber \\
&& \left. ~~~~
+~ \frac{6307462}{6561} \Nf^2 T_F^2 C_F^2
+ \frac{26893727}{1728} \zeta_3 C_F C_A^3
\right. \nonumber \\
&& \left. ~~~~
+~ \frac{164379529}{11664} \Nf T_F C_F C_A^2
+ \frac{1720647577}{104976} C_F^2 C_A^2
- 80703 \zeta_7 C_F^3 C_A
\right. \nonumber \\
&& \left. ~~~~
-~ 29080 \zeta_5 C_F^4
- 11066 \zeta_3^2 C_F^2 C_A^2
- 882 \zeta_7 \Nf T_F C_F C_A^2
\right. \nonumber \\
&& \left. ~~~~
+~ 992 \zeta_3^2 \Nf T_F C_F^2 C_A
+ 2352 \zeta_7 \Nf T_F C_F^2 C_A
\right] a^4 ~+~ O(a^5)
\end{eqnarray}
in the Landau gauge where $\Nc$ is the dimension of the quark representation. 
This together with all the other arbitrary gauge and colour group form factors 
and operator correlation functions are available in electronic form in the 
ancillary file accessible from the arXiv version of this article. The 
expressions for the remaining form factors are analytically similar and given 
in Appendix A. For practical purposes it is more appropriate to record the 
numerical values which for $SU(3)$ are
\begin{eqnarray} 
\left. \Sigma^{(1)}_1 (p) \right|_{p^2 = \mu^2}^{SU(3)} &=& 1 - 4.592593 a 
+ [ - 0.388889 \alpha^2 - 0.283122 \alpha + 8.646091 \Nf - 73.270703 ] a^2 
\nonumber \\
&&
+~ [ 461.135821 \Nf - 3.287545 \alpha^3 - 12.959452 \alpha^2 
- 5.608009 \alpha \Nf
\nonumber \\
&& ~~~~
+~ 39.666431 \alpha - 10.960314 \Nf^2 - 2351.601164 ] a^3 
\nonumber \\
&&
+~ [ 21.472542 \Nf^3 - 38.023637 \alpha^4 - 194.4294845 \alpha^3 
+ 22.752768 \alpha^2 \Nf 
\nonumber \\
&& ~~~~
-~ 540.676547 \alpha^2 - 4.0372388 \alpha \Nf^2 
+ 61.519123 \alpha \Nf - 615.036959 \alpha 
\nonumber \\
&& ~~~~
-~ 1774.989734 \Nf^2 
+ 27382.719248 \Nf - 97581.592304 ] a^4 ~+~ O(a^5) \nonumber \\
\left. \Sigma^{(2)}_1(p) \right|_{p^2 = \mu^2}^{SU(3)} &=& 
-~ [ 2.666667 \alpha + 0.888889 ] a \nonumber \\
&& +~ [ 4.444444 \Nf - 9.777778 \alpha^2 - 22.136631 \alpha - 87.712845 ] a^2 
\nonumber \\
&&
+~ [ 544.087233 \Nf - 59.648148 \alpha^3 - 226.804663 \alpha^2 
+ 32.281594 \alpha \Nf 
\nonumber \\
&& ~~~~
-~ 584.544161 \alpha - 13.278464 \Nf^2 
- 3888.145045 ] a^3 
\nonumber \\
&&
+~ [ 50259.534824 \Nf - 466.523059 \alpha^4 - 2370.497974 \alpha^3 
+ 416.103152 \alpha^2 \Nf 
\nonumber \\
&& ~~~~
-~ 8512.621550 \alpha^2 - 78.992915 \alpha \Nf^2 
+ 3621.132590 \alpha \Nf 
\nonumber \\
&& ~~~~
-~ 26911.574616 \alpha + 42.883707 \Nf^3 
- 2901.997856 \Nf^2 - 22151.706910 ] a^4 \nonumber \\
&& +~ O(a^5)
\end{eqnarray}
for a general linear covariant gauge. The parallel expressions for the other
non-zero independent form factors are 
\begin{eqnarray} 
\left. \widehat{\Sigma}^{(1)}_1(p) \right|_{p^2 = \mu^2}^{SU(3)} &=&
1 + [ 1.333333 \alpha - 4.740741 ] a \nonumber \\
&& +~ [ 4.500000 \alpha^2 + 5.254329 \alpha + 7.082304 \Nf - 68.174225 ] a^2
\nonumber \\
&& +~ [ 26.536529 \alpha^3 + 81.897981 \alpha^2 - 17.407350 \alpha \Nf
+ 199.116431 \alpha - 11.111158 \Nf^2
\nonumber \\
&& ~~~~
+~ 435.809244 \Nf - 2232.899286 ] a^3
\nonumber \\
&& +~ [ 195.237892 \alpha^4 + 897.152056 \alpha^3 - 170.720362 \alpha^2 \Nf 
+ 3037.517384 \alpha^2 
\nonumber \\
&& ~~~~
+~ 23.054518 \alpha \Nf^2 - 1135.908894 \alpha \Nf + 7925.459050 \alpha 
+ 25.898876 \Nf^3 
\nonumber \\
&& ~~~~
-~ 1832.438346 \Nf^2 + 27391.205548 \Nf - 100557.006490 ] a^4 ~+~ O(a^5) 
\nonumber \\
\left. \widehat{\Sigma}^{(2)}_1(p) \right|_{p^2 = \mu^2}^{SU(3)} &=& 
3.555556 a + [ 4.740741 \alpha - 5.925926 \Nf + 68.075731 ] a^2
\nonumber \\
&& +~ [ 16.000000 \alpha^2 - 7.901235 \alpha \Nf + 143.311350 \alpha
+ 13.607682 \Nf^2 
\nonumber \\
&& ~~~~
-~ 505.740039 \Nf + 2852.796746 ] a^3
\nonumber \\
&& +~ [ 94.352104 \alpha^3 - 26.666667 \alpha^2 \Nf + 720.298117 \alpha^2 
+ 18.143576 \alpha \Nf^2 
\nonumber \\
&& ~~~~
-~ 863.252767 \alpha \Nf + 6216.513064 \alpha - 35.458009 \Nf^3 
\nonumber \\
&& ~~~~
+~ 2344.739666 \Nf^2 - 36248.056464 \Nf + 145064.413688 ] a^4 ~+~ O(a^5)
\nonumber \\
\left. \widehat{\Sigma}^{(1)}_2(p) \right|_{p^2 = \mu^2}^{SU(3)} &=& 
1 + [ 1.333333 \alpha - 2.666667 ] a 
\nonumber \\
&&
+~ [ 4.500000 \alpha^2 + 8.019762 \alpha + 1.518519 \Nf - 25.808228 ] a^2 
\nonumber \\
&&
+~ [ 26.536529 \alpha^3 + 91.231315 \alpha^2 - 24.825732 \alpha \Nf 
\nonumber \\
&& ~~~~
+~ 277.631323 \alpha - 1.76955 \Nf^2 + 104.535817 \Nf - 567.226623 ] a^3 
\nonumber \\
&&
+~ [ 195.237892 \alpha^4 + 952.190783 \alpha^3 - 195.757399 \alpha^2 \Nf 
+ 3432.896937 \alpha^2 
\nonumber \\
&& ~~~~
+~ 35.509999 \alpha \Nf^2 - 1694.463551 \alpha \Nf 
+ 11252.330095 \alpha + 4.762033 \Nf^3 
\nonumber \\
&& ~~~~
-~ 346.382688 \Nf^2 + 5695.927082 \Nf 
- 22634.2729429 ] a^4 ~+~ O(a^5) \nonumber \\
\left. \widehat{\Sigma}^{(2)}_2(p) \right|_{p^2 = \mu^2}^{SU(3)} &=& 
5.333333 a + [ 7.111111 \alpha - 7.703704 \Nf + 102.544867 ] a^2 
\nonumber \\
&&
+~ [ 24.000000 \alpha^2 - 10.271605 \alpha \Nf + 215.310880 \alpha 
\nonumber \\
&& ~~~~
+~ 16.460905 \Nf^2 - 701.956934 \Nf + 4114.414245 ] a^3 
\nonumber \\
&&
+~ [ 141.528156 \alpha^3 - 34.666667 \alpha^2 \Nf + 1076.563173 \alpha^2 
+ 21.947874 \alpha \Nf^2 
\nonumber \\
&& ~~~~
-~ 1200.648970 \alpha \Nf + 9112.572345 \alpha 
- 39.579332 \Nf^3 + 2938.777905 \Nf^2 
\nonumber \\
&& ~~~~
-~ 48591.964797 \Nf + 2014630.937750 ] a^4 ~+~ O(a^5) \nonumber \\
\left. \widetilde{\Sigma}^{(1)}_1(p) \right|_{p^2 = \mu^2}^{SU(3)} &=&
-~ 2.074074 a 
+ [ 5.563786 \Nf - 2.765432 \alpha - 42.366000 ] a^2 
\nonumber \\
&& +~ [ 7.418381 \alpha \Nf - 9.333333 \alpha^2 - 78.514892 \alpha 
- 9.341611 \Nf^2 
\nonumber \\
&& ~~~~
+~ 331.273427 \Nf - 1665.672663 ] a^3 
\nonumber \\
&& +~ [ 25.037037 \alpha^2 \Nf - 55.038727 \alpha^3 - 395.379553 \alpha^2 
- 12.455481 \alpha \Nf^2 
\nonumber \\
&& ~~~~
+~ 558.554656 \alpha \Nf - 3326.871045 \alpha + 21.136843 \Nf^3 
- 1486.055658 \Nf^2 
\nonumber \\
&& ~~~~
+~ 21695.278466 \Nf - 77922.733548 ] a^4 ~+~ O(a^5) \nonumber \\
\left. \widetilde{\Sigma}^{(2)}_1(p) \right|_{p^2 = \mu^2}^{SU(3)} &=&
-~ 1.777778 a 
+ [ 1.777778 \Nf - 2.370370 \alpha - 34.469136 ] a^2 
\nonumber \\
&& +~ [ 2.370370 \alpha \Nf - 8.000000 \alpha^2 - 71.999531 \alpha 
- 2.853224 \Nf^2 
\nonumber \\
&& ~~~~
+~ 196.216895 \Nf - 1261.617500 ] a^3 
\nonumber \\
&& +~ [ 8.000000 \alpha^2 \Nf - 47.176052 \alpha^3 - 356.265056 \alpha^2 
- 3.804298 \alpha \Nf^2 
\nonumber \\
&& ~~~~
+~ 337.396203 \alpha \Nf - 2896.059280 \alpha + 4.121323 \Nf^3 
- 594.038239 \Nf^2 
\nonumber \\
&& ~~~~
+~ 12343.908333 \Nf - 56398.680086 ] a^4 ~+~ O(a^5) 
\end{eqnarray}
for an arbitrary number of quarks.

{\begin{figure}[ht]
\begin{center}
\includegraphics[width=8.0cm,height=4.0cm]{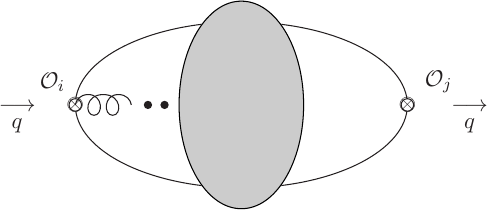}
\end{center}
\caption{Operator correlation function 
$\Pi_{ij}^{\mu_1 \mu_2 \nu_1 \nu_2}(q)$.}
\label{figopcor}
\end{figure}}

\sect{Operator correlation functions.}

The second class of Green's function that we examine is that concerning the
correlation function of the two gauge invariant operators ${\cal O}_i^{\mu\nu}$
defined as
\begin{equation}
\Pi_{ij}^{\mu_1 \mu_2 \nu_1 \nu_2}(q) ~=~ (4\pi)^2 i
\int \, d^d x \, e^{iqx} \langle 0 | {\cal O}_i^{\mu_1 \mu_2}(x)
{\cal O}_j^{\nu_1 \nu_2}(0) | 0 \rangle
\label{opcordef}
\end{equation}
where $q$ is the momentum with $q^2$~$=$~$-$~$Q^2$. The correlation function is
formally illustrated in Figure \ref{figopcor} where the partially attached
gluon at ${\cal O}_i$ notionally represents one of the two ${\cal O}_1$ 
operator Feynman rules while operator ${\cal O}_j$ indicates the other 
eventuality for both ${\cal O}_1$ and ${\cal O}_2$. Then the four combinations 
at both insertions in Figure \ref{figopcor} cover all possible Feynman graph 
structures of (\ref{opcordef}). We will follow the same notation and 
conventions as the three loop computation of \cite{34}. It is worth recalling 
that in this section the loop order does not equate to the power of $a$ in the 
perturbative expansion of the correlation function unlike the previous
section. For instance, the leading graph of Figure \ref{figopcor} is one loop 
but as it does not contain any interactions it is therefore $O(1)$ and not 
$O(a)$. Since both operators are gauge invariant this implies that 
$\Pi_{ij}^{\mu_1 \mu_2 \nu_1 \nu_2}(q)$ is independent of the covariant gauge 
parameter in the $\MSbar$ scheme which we will use for our computation. The 
mixing of the operators will be treated in the same way as in the previous 
section. By examining the possible rank four Lorentz tensors built from the 
metric $\eta_{\mu\nu}$ and the external momentum $q_\mu$ and recalling the 
symmetric and tracelessness property of the two twist-$2$ operators there are 
three independent tensors into which the correlation function can be 
decomposed. In other words we can rewrite (\ref{opcordef}) as
\begin{equation}
\Pi_{ij}^{\mu_1 \mu_2 \nu_1 \nu_2}(q) ~=~
\sum_{k=1}^3 {\cal P}_{(k)}^{\{ \mu_1 \mu_2 | \nu_1 \nu_2 \} }(q) \, 
\Pi_{{ij}\,(k)}(q^2)
\end{equation}
where 
\begin{eqnarray}
{\cal P}_{(1)}^{ \{\mu \nu | \sigma \rho \} }(q) &=&
\eta^{\mu\sigma} \eta^{\nu\rho} ~+~ \eta^{\mu\rho} \eta^{\nu\sigma} ~-~
\frac{2}{d} \eta^{\mu\nu} \eta^{\sigma\rho} \nonumber \\
{\cal P}_{(2)}^{ \{\mu \nu | \sigma \rho \} }(q) &=&
-~ \frac{1}{d} \eta^{\mu\nu} \eta^{\sigma\rho} ~+~ \left[ \eta^{\mu\nu}
q^\sigma q^\rho ~+~ \eta^{\sigma\rho} q^\mu q^\nu \right] \frac{1}{q^2} ~-~
d \frac{q^\mu q^\nu q^\sigma q^\rho}{(q^2)^2} \nonumber \\
{\cal P}_{(3)}^{ \{\mu \nu | \sigma \rho \} }(q) &=&
P^{\mu\sigma}(q) L^{\nu\rho}(q) ~+~ P^{\mu\rho}(q) L^{\nu\sigma}(q) ~+~
P^{\nu\sigma}(q) L^{\mu\rho}(q) ~+~ P^{\nu\rho}(q) L^{\mu\sigma}(q) 
\end{eqnarray}
with
\begin{equation}
P_{\mu\nu}(q) ~=~ \eta_{\mu\nu} ~-~ \frac{q_\mu q_\nu}{q^2} ~~~,~~~
L_{\mu\nu}(q) ~=~ \frac{q_\mu q_\nu}{q^2} 
\end{equation}
and $\Pi_{{ij}\,(k)}(q^2)$ are the scalar functions we will determine to four
loops. The projections required to extract these are given by
\begin{equation}
\Pi_{{ij}\,(k)}(q^2) ~=~ \sum_{l=1}^3 {\cal M}_{kl}
{\cal P}_{(l)}^{\{ \mu \nu | \rho \sigma \} }(q) 
\Pi_{ij \, \mu \nu \rho \sigma}(q)
\end{equation}
for each value of $k$ where the matrix ${\cal M}$ is formally deduced from
\begin{equation}
{\cal M}^{-1}_{ij} ~=~
{\cal P}_{(i)}^{\{ \mu \nu | \rho \sigma \} }(q) 
{\cal P}_{(j)\,\{ \mu \nu | \rho \sigma \} }(q) ~.
\end{equation}
In particular a straightforward computation gives
\begin{equation}
{\cal M} ~=~ \frac{1}{4(d-1)(d+1)(d-2)(q^2)^2}
\left(
\begin{array}{ccc}
2 (d-1) & 4 & -~ 2 (d-1) \\
4 & 4 d & -~ 4 \\
-~ 2 (d-1) & -~ 4 & (d^2+d-4) \\
\end{array}
\right) ~.
\end{equation}
Implementing this projection on the correlation functions determines the scalar
form factors which can then be passed to the {\sc Forcer} algorithm which will
evaluate the contributing Feynman graphs. These are generated by {\sc Qgraf},
\cite{36}, with the number of graphs for each of the three Green's function we 
will determine indicated in Table \ref{opcornum}.

{\begin{table}[ht]
\begin{center}
\begin{tabular}{|c||r|r|r|}
\hline
\rule{0pt}{12pt}
$L$ & $\Pi_{11}(q^2)$ & $\Pi_{12}(q^2)$ & $\Pi_{22}(q^2)$ \\
\hline
$1$ & $1$ & $1$ & $1$ \\
$2$ & $8$ & $5$ & $3$ \\
$3$ & $92$ & $36$ & $30$ \\
$4$ & $1560$ & $871$ & $481$ \\
\hline
Total & $1661$ & $913$ & $518$ \\
\hline
\end{tabular}
\end{center}
\begin{center}
\caption{Number of graphs at each loop order $L$ for operator correlation
functions of (\ref{opcordef}).}
\label{opcornum}
\end{center}
\end{table}}

In the evaluation of the correlation functions several aspects of their
renormalization have to be accounted for. First as these involve operators
rather than fields the operator mixing has to be included. At the next level
the correlation functions are not multiplicatively renormalizable. Instead 
the correlators involve a contact term, \cite{39,40}, meaning each of the three
Lorentz projections require an associated renormalization constant. For
operator correlations where there is no operator mixing this means an
additional renormalization constant with a corresponding renormalization
group function. In the case of the twist-$2$ operators it is appropriate to be
specific and note the relation of the renormalized correlation function form
factors to their bare counterparts are, \cite{34},
\begin{eqnarray}
\Pi_{11\,(i)}(q) &=& (q^2)^2 Z_{11\,(i)} ~+~ 
\mu^{2\epsilon} \left[ Z_{11}^2
\Pi_{11\,(i)\,\mbox{\footnotesize{o}}}(q) ~+~
2 Z_{11} Z_{12}
\Pi_{12\,(i)\,\mbox{\footnotesize{o}}}(q) ~+~
Z_{12}^2 \Pi_{22\,(i)\,\mbox{\footnotesize{o}}}(q) \right]
\nonumber \\
\Pi_{12\,(i)}(q) &=& (q^2)^2 Z_{12\,(i)} ~+~ 
\mu^{2\epsilon} \left[ Z_{11} Z_{22}
\Pi_{12\,(i)\,\mbox{\footnotesize{o}}}(q) ~+~
Z_{12} Z_{22}
\Pi_{22\,(i)\,\mbox{\footnotesize{o}}}(q) \right]
\nonumber \\
\Pi_{22\,(i)}(q) &=& (q^2)^2 Z_{22\,(i)} ~+~ \mu^{2\epsilon} Z_{22}^2
\Pi_{22\,(i)\,\mbox{\footnotesize{o}}}(q) 
\end{eqnarray}
where $Z_{ij\,(k)}$ are the contact renormalization constants which are $O(1)$,
and start with a counterterm and not unity, and $Z_{ij}$ is the mixing matrix 
for the operator renormalization given in (\ref{opmixmat}). It is 
straightforward to deduce that the renormalized correlator form factors 
$\Pi_{ij\,(k)}$ satisfy
\begin{eqnarray}
0 &=& \mu \frac{d~}{d\mu} \Pi_{11\,(i)}(q) ~+~ 
2 \gamma_{11}(a) \Pi_{11\,(i)}(q) ~+~ 2 \gamma_{12}(a) \Pi_{12\,(i)}(q) ~-~ 
(q^2)^2 \gamma_{11\,(i)}(a) \nonumber \\
0 &=& \mu \frac{d~}{d\mu} \Pi_{12\,(i)}(q) ~+~ 
\left[ \gamma_{11}(a) + \gamma_{22}(a) \right] \Pi_{12\,(i)}(q) ~+~
\gamma_{12}(a) \Pi_{22\,(i)}(q) ~-~ (q^2)^2 \gamma_{12\,(i)}(a) \nonumber \\
0 &=& \mu \frac{d~}{d\mu} \Pi_{22\,(i)}(q) ~+~ 
2 \gamma_{22}(a) \Pi_{22\,(i)}(q) ~-~ (q^2)^2 \gamma_{22\,(i)}(a) ~.
\end{eqnarray}
The respective contact anomalous dimensions, $\gamma_{ij\,(k)}(a)$, are deduced
from $Z_{ij\,(k)}$ from the relations
\begin{eqnarray}
\gamma_{11\,(i)}(a) &=& \left[ -~ \epsilon ~+~ \beta(a) 
\frac{\partial~}{\partial a} ~+~ 2 \gamma_{11}(a) 
\right] Z_{11\,(i)} ~+~ 2 \gamma_{12}(a) Z_{12\,(i)} \nonumber \\
\gamma_{12\,(i)}(a) &=& \left[ -~ \epsilon ~+~ \beta(a) 
\frac{\partial~}{\partial a} ~+~ \gamma_{11}(a) ~+~ \gamma_{22}(a) 
\right] Z_{12\,(i)} ~+~ \gamma_{12}(a) Z_{22\,(i)} \nonumber \\
\gamma_{22\,(i)}(a) &=& \left[ -~ \epsilon ~+~ 
\beta(a) \frac{\partial~}{\partial a} ~+~ 2 \gamma_{22}(a) \right] Z_{22\,(i)}
\end{eqnarray} 
where we have assumed there is no gauge parameter dependence in the application
of (\ref{muderive}) as we only consider the $\MSbar$ scheme. We note that
$\gamma_{ij}(a)$ are given by (\ref{gamma11}), (\ref{gamma12}) and
\ref{gamma22}) although only the three loop expressions are needed to deduce
$\gamma_{ij\,(k)}(a)$ to four loops which equates to $O(a^3)$ for 
(\ref{opcordef}).

Applying the {\sc Forcer} algorithm followed by the renormalization formalism 
we arrive at the four loop contact renormalization group functions which are 
\begin{eqnarray}
\gamma_{11\,(1)}(a) &=& \left[
\frac{1}{5}
+ \frac{103}{225} C_F a
\right. \nonumber \\
&& \left. ~
+ \left[
\frac{5653}{10125} C_F T_F \Nf
+ \frac{162749}{20250} C_F^2
- \frac{65603}{40500} C_A C_F
- \frac{64}{5} \zeta_3 C_F^2
+ \frac{32}{5} \zeta_3 C_A C_F
\right] a^2 \right. \nonumber \\
&& \left. ~+ \left[
128 \zeta_5 C_F^3
- \frac{93921197}{5467500} C_A^2 C_F
- \frac{7621891}{1366875} C_A C_F T_F \Nf
- \frac{5319308}{1366875} C_F^2 T_F \Nf
\right. \right. \nonumber \\
&& \left. \left. ~~~~~
- \frac{130504}{675} \zeta_3 C_A C_F^2
- \frac{44576}{675} \zeta_3 C_F^3
- \frac{14896}{675} \zeta_3 C_A C_F T_F \Nf
- \frac{512}{135} \zeta_3 C_F T_F^2 \Nf^2
\right. \right. \nonumber \\
&& \left. \left. ~~~~~
- \frac{416}{15} \zeta_4 C_A^2 C_F
- \frac{128}{3} \zeta_4 C_F^3
- \frac{128}{5} \zeta_4 C_F^2 T_F \Nf
+ \frac{64}{3} \zeta_5 C_A C_F^2
+ \frac{384}{5} \zeta_4 C_A C_F^2
\right. \right. \nonumber \\
&& \left. \left. ~~~~~
+ \frac{256}{15} \zeta_4 C_A C_F T_F \Nf
+ \frac{688}{15} \zeta_3 C_F^2 T_F \Nf
+ \frac{1784}{15} \zeta_3 C_A^2 C_F
+ \frac{4267972}{1366875} C_F T_F^2 \Nf^2
\right. \right. \nonumber \\
&& \left. \left. ~~~~~
+ \frac{9163637}{2733750} C_F^3
+ \frac{110800357}{1366875} C_A C_F^2
- 64 \zeta_5 C_A^2 C_F
\right] a^3 \right] \Nf \Nc ~+~ O(a^4) \nonumber \\
\gamma_{11\,(2)}(a) &=& \left[
\frac{2}{15}
+ \frac{18}{25} C_F a
\right. \nonumber \\
&& \left. ~
+ \left[
\frac{64}{15} \zeta_3 C_A C_F
+ \frac{26507}{60750} C_A C_F
+ \frac{57623}{10125} C_F^2
- \frac{3914}{30375} C_F T_F \Nf
- \frac{128}{15} \zeta_3 C_F^2
\right] a^2 \right. \nonumber \\
&& \left. ~+ \left[
\frac{128}{9} \zeta_5 C_A C_F^2
- \frac{9306314}{1366875} C_A C_F T_F \Nf
- \frac{6748732}{1366875} C_F^2 T_F \Nf
- \frac{91072}{2025} \zeta_3 C_F^3
\right. \right. \nonumber \\
&& \left. \left. ~~~~~
- \frac{38752}{2025} \zeta_3 C_A C_F T_F \Nf
- \frac{9104}{75} \zeta_3 C_A C_F^2
- \frac{1024}{405} \zeta_3 C_F T_F^2 \Nf^2
- \frac{832}{45} \zeta_4 C_A^2 C_F
\right. \right. \nonumber \\
&& \left. \left. ~~~~~
- \frac{256}{9} \zeta_4 C_F^3
- \frac{256}{15} \zeta_4 C_F^2 T_F \Nf
- \frac{128}{3} \zeta_5 C_A^2 C_F
+ \frac{256}{3} \zeta_5 C_F^3
+ \frac{256}{5} \zeta_4 C_A C_F^2
\right. \right. \nonumber \\
&& \left. \left. ~~~~~
+ \frac{512}{45} \zeta_4 C_A C_F T_F \Nf
+ \frac{4384}{135} \zeta_3 C_F^2 T_F \Nf
+ \frac{30992}{405} \zeta_3 C_A^2 C_F
+ \frac{303803}{1366875} C_A^2 C_F
\right. \right. \nonumber \\
&& \left. \left. ~~~~~
+ \frac{1969633}{455625} C_F^3
+ \frac{2141888}{1366875} C_F T_F^2 \Nf^2
+ \frac{61801178}{1366875} C_A C_F^2
\right] a^3 \right] \Nf \Nc ~+~ O(a^4) \nonumber \\
\gamma_{11\,(3)}(a) &=& \left[
\frac{2}{15}
+ \frac{62}{225} C_F a
\right. \nonumber \\
&& \left. ~
+ \left[
\frac{64}{5} \zeta_3 C_F^2
+ \frac{78919}{20250} C_A C_F
- \frac{92377}{10125} C_F^2
- \frac{13138}{10125} C_F T_F \Nf
- \frac{32}{5} \zeta_3 C_A C_F
\right] a^2 \right. \nonumber \\
&& + \left. \left[
64 \zeta_5 C_A^2 C_F
- \frac{92115697}{1366875} C_A C_F^2
- \frac{31604851}{1366875} C_F^3
- \frac{5875612}{1366875} C_F T_F^2 \Nf^2
\right. \right. \nonumber \\
&& \left. \left. ~~~~~
- \frac{4596232}{1366875} C_F^2 T_F \Nf
- \frac{4832}{45} \zeta_3 C_A^2 C_F
- \frac{1312}{45} \zeta_3 C_F^2 T_F \Nf
- \frac{384}{5} \zeta_4 C_A C_F^2
\right. \right. \nonumber \\
&& \left. \left. ~~~~~
- \frac{256}{15} \zeta_4 C_A C_F T_F \Nf
- \frac{64}{3} \zeta_5 C_A C_F^2
+ \frac{128}{3} \zeta_4 C_F^3
+ \frac{128}{5} \zeta_4 C_F^2 T_F \Nf
\right. \right. \nonumber \\
&& \left. \left. ~~~~~
+ \frac{416}{15} \zeta_4 C_A^2 C_F
+ \frac{512}{135} \zeta_3 C_F T_F^2 \Nf^2
+ \frac{3616}{675} \zeta_3 C_A C_F T_F \Nf
+ \frac{49376}{675} \zeta_3 C_F^3
\right. \right. \nonumber \\
&& \left. \left. ~~~~~
+ \frac{123664}{675} \zeta_3 C_A C_F^2
+ \frac{8532436}{1366875} C_A C_F T_F \Nf
+ \frac{72395837}{5467500} C_A^2 C_F
\right. \right. \nonumber \\
&& \left. \left. ~~~~~
- 128 \zeta_5 C_F^3
\right] a^3 \right] \Nf \Nc ~+~ O(a^4) \nonumber \\
\gamma_{12\,(1)}(a) &=& O(a^4) ~~~,~~~ 
\gamma_{12\,(2)}(a) ~=~ O(a^4) \nonumber \\
\gamma_{12\,(3)}(a) &=& \left[
\frac{2}{3}
+ 2 C_F a
+ \left[
\frac{133}{18} C_A C_F
- \frac{22}{9} C_F T_F \Nf
- C_F^2
\right] a^2 \right. \nonumber \\
&& \left. ~+ \left[
\frac{176}{9} \zeta_3 C_A^2 C_F
+ \frac{352}{9} \zeta_3 C_F^2 T_F \Nf
+ \frac{860}{27} C_A C_F^2
+ \frac{5815}{486} C_A^2 C_F
\right. \right. \nonumber \\
&& \left. \left. ~~~~~
- \frac{1538}{243} C_A C_F T_F \Nf
- \frac{676}{27} C_F^2 T_F \Nf
- \frac{616}{243} C_F T_F^2 \Nf^2
- \frac{352}{9} \zeta_3 C_A C_F T_F \Nf
\right. \right. \nonumber \\
&& \left. \left. ~~~~~
- \frac{176}{9} \zeta_3 C_A C_F^2
- 23 C_F^3
\right] a^3 \right] \Nf \Nc ~+~ O(a^4) \nonumber \\
\gamma_{22\,(1)}(a) &=& O(a^4) ~~~,~~~ 
\gamma_{22\,(2)}(a) ~=~ O(a^4) ~~~,~~~ 
\gamma_{22\,(3)}(a) ~=~ 2 \gamma_{12\,(3)}(a) ~+~ O(a^4) ~.
\label{opcoranom}
\end{eqnarray}
We have reproduced the previous three loop expressions, \cite{34}, determined
with the {\sc Mincer} package which is a partial check on our {\sc Forcer}
computation. In carrying out the calculation for a non-zero gauge parameter 
$\alpha$ one key check is that the expressions are independent of $\alpha$. One
final verification of the explicit computation is that we checked the final 
three relations of (\ref{opcoranom}) were satisfied to $O(a^3)$.

To complete the four loop properties we record the expressions for the form 
factors. First factoring off the momentum scale via
\begin{equation}
\Pi_{ij\,(k)}(a) ~=~ (q^2)^2 \tilde{\Pi}_{ij\,(k)}(a) 
\end{equation} 
we have 
\begin{eqnarray}
\tilde{\Pi}_{11\,(1)}(a) &=&
\left[ 
\frac{12}{25}
- \left[
\frac{17533}{13500} C_F
+ \frac{12}{5} \zeta_3 C_F
\right] a \right. \nonumber \\
&& \left.
+ \left[
\frac{16}{5} \zeta_4 C_F C_A
- \frac{3541817}{1215000} C_F C_A
- \frac{15838}{675} \zeta_3 C_F C_A
- \frac{2606}{135} \zeta_3 C_F^2
- 4 \zeta_5 C_F C_A
\right. \right. \nonumber \\
&& \left. \left. ~~~~
+~ \frac{1816}{135} \zeta_3 \Nf T_F C_F
+ \frac{419327}{303750} \Nf T_F C_F
+ \frac{12235087}{455625} C_F^2
- \frac{32}{5} \zeta_4 C_F^2
+ 24 \zeta_5 C_F^2
\right] a^2 \right. \nonumber \\
&& \left.
+ \left[
\frac{14603329771}{41006250} C_F^2 C_A
+ \frac{71168258749}{164025000} C_F C_A^2
- \frac{30822165083}{164025000} C_F^3
\right. \right. \nonumber \\
&& \left. \left. ~~~~
-~ \frac{29868617111}{82012500} \Nf T_F C_F C_A
- \frac{4722649781}{164025000} \Nf T_F C_F^2
- \frac{13558331}{20250} \zeta_3 C_F^2 C_A
\right. \right. \nonumber \\
&& \left. \left. ~~~~
-~ \frac{9337969}{30375} \zeta_3 C_F C_A^2
- \frac{3283718}{30375} \zeta_3 C_F^3
- \frac{546728}{6075} \zeta_3 \Nf^2 T_F^2 C_F
- \frac{16909}{30} \zeta_5 C_F C_A^2
\right. \right. \nonumber \\
&& \left. \left. ~~~~
-~ \frac{9401}{225} \zeta_4 C_F^2 C_A
- \frac{2036}{45} \zeta_4 C_F^3
- \frac{1696}{5} \zeta_5 \Nf T_F C_F^2
- \frac{192}{5} \zeta_3^2 \Nf T_F C_F^2
\right. \right. \nonumber \\
&& \left. \left. ~~~~
-~ \frac{82}{25} \zeta_4 \Nf T_F C_F^2
- \frac{64}{3} \zeta_5 \Nf^2 T_F^2 C_F
- \frac{64}{45} \zeta_4 \Nf^2 T_F^2 C_F
+ \frac{32}{5} \zeta_3^2 \Nf T_F C_F C_A
\right. \right. \nonumber \\
&& \left. \left. ~~~~
+~ \frac{40}{3} \zeta_6 C_F^2 C_A
+ \frac{264}{5} \zeta_3^2 C_F C_A^2
+ \frac{272}{15} \zeta_3^2 C_F^2 C_A
+ \frac{859}{25} \zeta_4 C_F C_A^2
\right. \right. \nonumber \\
&& \left. \left. ~~~~
+~ \frac{2746}{225} \zeta_4 \Nf T_F C_F C_A
+ \frac{7807}{9} \zeta_5 C_F^2 C_A
+ \frac{10096}{45} \zeta_5 \Nf T_F C_F C_A
+ \frac{15502}{45} \zeta_5 C_F^3
\right. \right. \nonumber \\
&& \left. \left. ~~~~
+~ \frac{631028}{3375} \zeta_3 \Nf T_F C_F^2
+ \frac{13729699}{30375} \zeta_3 \Nf T_F C_F C_A
+ \frac{1415796323}{20503125} \Nf^2 T_F^2 C_F
\right. \right. \nonumber \\
&& \left. \left. ~~~~
-~ 336 \zeta_7 C_F^3
- 40 \zeta_6 C_F C_A^2
- 28 \zeta_7 C_F C_A^2
+ 32 \zeta_3^2 C_F^3
+ 80 \zeta_6 C_F^3
\right. \right. \nonumber \\
&& \left. \left. ~~~~
+~ 168 \zeta_7 C_F^2 C_A
\right] a^3
\right] \Nc ~+~ O(a^4) \nonumber \\
\tilde{\Pi}_{11\,(2)}(a) &=&
\left[ 
\frac{92}{225}
- \left[
\frac{15683}{20250} C_F
+ \frac{8}{5} \zeta_3 C_F
\right] a \right. \nonumber \\
&& \left.
+ \left[
\frac{32}{15} \zeta_4 C_F C_A
- \frac{7819793}{2733750} C_F C_A
- \frac{29276}{2025} \zeta_3 C_F C_A
- \frac{5788}{405} \zeta_3 C_F^2
- \frac{8}{3} \zeta_5 C_F C_A
\right. \right. \nonumber \\
&& \left. \left. ~~~~
-~ \frac{64}{15} \zeta_4 C_F^2
+ \frac{3632}{405} \zeta_3 \Nf T_F C_F
+ \frac{1685066}{1366875} \Nf T_F C_F
+ \frac{8792588}{455625} C_F^2
+ 16 \zeta_5 C_F^2
\right] a^2 \right. \nonumber \\
&& \left.
+ \left[
\frac{5674731737}{123018750} \Nf^2 T_F^2 C_F
+ \frac{29846417687}{123018750} C_F^2 C_A
+ \frac{544884357737}{1968300000} C_F C_A^2
\right. \right. \nonumber \\
&& \left. \left. ~~~~
-~ \frac{29359415471}{123018750} \Nf T_F C_F C_A
- \frac{15902453419}{123018750} C_F^3
- \frac{1003879454}{61509375} \Nf T_F C_F^2
\right. \right. \nonumber \\
&& \left. \left. ~~~~
-~ \frac{13904861}{30375} \zeta_3 C_F^2 C_A
- \frac{5550136}{30375} \zeta_3 C_F C_A^2
- \frac{2315152}{30375} \zeta_3 C_F^3
\right. \right. \nonumber \\
&& \left. \left. ~~~~
-~ \frac{122384}{2025} \zeta_3 \Nf^2 T_F^2 C_F
- \frac{5783}{15} \zeta_5 C_F C_A^2
- \frac{4354}{225} \zeta_4 C_F^2 C_A
- \frac{3556}{675} \zeta_4 \Nf T_F C_F^2
\right. \right. \nonumber \\
&& \left. \left. ~~~~
-~ \frac{3392}{15} \zeta_5 \Nf T_F C_F^2
- \frac{1544}{45} \zeta_4 C_F^3
- \frac{128}{5} \zeta_3^2 \Nf T_F C_F^2
- \frac{128}{9} \zeta_5 \Nf^2 T_F^2 C_F
\right. \right. \nonumber \\
&& \left. \left. ~~~~
-~ \frac{128}{135} \zeta_4 \Nf^2 T_F^2 C_F
- \frac{80}{3} \zeta_6 C_F C_A^2
- \frac{56}{3} \zeta_7 C_F C_A^2
+ \frac{64}{3} \zeta_3^2 C_F^3
+ \frac{64}{15} \zeta_3^2 \Nf T_F C_F C_A
\right. \right. \nonumber \\
&& \left. \left. ~~~~
+ \frac{80}{9} \zeta_6 C_F^2 C_A
+~ \frac{160}{3} \zeta_6 C_F^3
+ \frac{176}{5} \zeta_3^2 C_F C_A^2
+ \frac{544}{45} \zeta_3^2 C_F^2 C_A
+ \frac{1498}{75} \zeta_4 C_F C_A^2
\right. \right. \nonumber \\
&& \left. \left. ~~~~
+~ \frac{1750}{3} \zeta_5 C_F^2 C_A
+ \frac{6932}{675} \zeta_4 \Nf T_F C_F C_A
+ \frac{20192}{135} \zeta_5 \Nf T_F C_F C_A
+ \frac{32204}{135} \zeta_5 C_F^3
\right. \right. \nonumber \\
&& \left. \left. ~~~~
+~ 112 \zeta_7 C_F^2 C_A
+ \frac{1252696}{10125} \zeta_3 \Nf T_F C_F^2
+ \frac{9021146}{30375} \zeta_3 \Nf T_F C_F C_A
\right. \right. \nonumber \\
&& \left. \left. ~~~~
-~ 224 \zeta_7 C_F^3
\right]
\right] \Nc ~+~ O(a^4) \nonumber \\
\tilde{\Pi}_{11\,(3)}(a) &=&
\left[ 
\frac{17}{225}
+ \left[
\frac{34439}{6750} C_F
- \frac{8}{5} \zeta_3 C_F
\right] a \right. \nonumber \\
&& \left.
+ \left[
\frac{32}{5} \zeta_4 C_F^2
+ \frac{464}{135} \zeta_3 \Nf T_F C_F
+ \frac{918157}{15000} C_F C_A
- \frac{8}{3} \zeta_5 C_F C_A
- \frac{28297949}{911250} C_F^2
\right. \right. \nonumber \\
&& \left. \left. ~~~~
-~ \frac{6263527}{303750} \Nf T_F C_F
- \frac{17492}{675} \zeta_3 C_F C_A
- \frac{436}{135} \zeta_3 C_F^2
- \frac{16}{5} \zeta_4 C_F C_A
+ 16 \zeta_5 C_F^2
\right] a^2 \right. \nonumber \\
&& \left.
+ \left[
\frac{4036036783}{82012500} \Nf^2 T_F^2 C_F
+ \frac{11393571679}{82012500} C_F^3
+ \frac{1142859633283}{1312200000} C_F C_A^2
\right. \right. \nonumber \\
&& \left. \left. ~~~~
-~ \frac{91459458709}{164025000} C_F^2 C_A
- \frac{19488148907}{41006250} \Nf T_F C_F C_A
- \frac{17002136}{30375} \zeta_3 C_F C_A^2
\right. \right. \nonumber \\
&& \left. \left. ~~~~
-~ \frac{1359662}{10125} \zeta_3 C_F^2 C_A
- \frac{4736}{45} \zeta_5 \Nf T_F C_F^2
- \frac{4156}{225} \zeta_4 \Nf T_F C_F C_A
- \frac{2252}{75} \zeta_4 C_F C_A^2
\right. \right. \nonumber \\
&& \left. \left. ~~~~
-~ \frac{272}{15} \zeta_3^2 C_F^2 C_A
- \frac{128}{5} \zeta_3^2 \Nf T_F C_F^2
- \frac{128}{9} \zeta_5 \Nf^2 T_F^2 C_F
- \frac{56}{3} \zeta_7 C_F C_A^2
\right. \right. \nonumber \\
&& \left. \left. ~~~~
-~ \frac{40}{3} \zeta_6 C_F^2 C_A
+ \frac{64}{15} \zeta_3^2 \Nf T_F C_F C_A
+ \frac{64}{45} \zeta_4 \Nf^2 T_F^2 C_F
+ \frac{716}{75} \zeta_4 \Nf T_F C_F^2
\right. \right. \nonumber \\
&& \left. \left. ~~~~
+~ \frac{1408}{15} \zeta_3^2 C_F C_A^2
+ \frac{1606}{9} \zeta_5 C_F^2 C_A
+ \frac{2156}{45} \zeta_4 C_F^3
+ \frac{3733}{135} \zeta_5 C_F C_A^2
+ \frac{5468}{45} \zeta_5 C_F^3
\right. \right. \nonumber \\
&& \left. \left. ~~~~
+~ \frac{8546}{225} \zeta_4 C_F^2 C_A
+ \frac{10912}{135} \zeta_5 \Nf T_F C_F C_A
+ \frac{60848}{6075} \zeta_3 \Nf^2 T_F^2 C_F
\right. \right. \nonumber \\
&& \left. \left. ~~~~
+~ \frac{265244}{1125} \zeta_3 \Nf T_F C_F^2
+ \frac{1782146}{30375} \zeta_3 \Nf T_F C_F C_A
+ \frac{4363598}{30375} \zeta_3 C_F^3
\right. \right. \nonumber \\
&& \left. \left. ~~~~
+~ \frac{971499581}{164025000} \Nf T_F C_F^2
- 224 \zeta_7 C_F^3
- 80 \zeta_6 C_F^3
- 32 \zeta_3^2 C_F^3
+ 40 \zeta_6 C_F C_A^2
\right. \right. \nonumber \\
&& \left. \left. ~~~~
+~ 112 \zeta_7 C_F^2 C_A
\right] a^3 
\right] \Nc ~+~ O(a^4) 
\end{eqnarray}
for the main operator ${\cal O}^{\mu\nu}_1$. For the remaining two
Green's functions only channel $3$ of the Lorentz decomposition is non-zero
since we found
\begin{eqnarray}
\tilde{\Pi}_{12\,(1)}(a) &=& 
\tilde{\Pi}_{12\,(2)}(a) ~=~ O(a^4) \nonumber \\
\tilde{\Pi}_{12\,(3)}(a) &=&
\left[
\frac{10}{9}
+ \left[
\frac{55}{6} C_F
- 8 \zeta_3 C_F
 \right] a \right. \nonumber \\
&& \left. 
+ \left[
\frac{304}{9} \zeta_3 \Nf T_F C_F
+ \frac{44215}{324} C_F C_A
+ 80 \zeta_5 C_F^2
- \frac{3701}{81} \Nf T_F C_F
- \frac{908}{9} \zeta_3 C_F C_A
\right. \right. \nonumber \\
&& \left. \left. ~~~~
-~ \frac{148}{3} \zeta_3 C_F^2
- \frac{143}{9} C_F^2
- \frac{40}{3} \zeta_5 C_F C_A
\right] a^2 \right. \nonumber \\
&& \left. 
+ \left[
52 \zeta_3 C_F^3
+ 560 \zeta_7 C_F^2 C_A
+ 980 \zeta_5 C_F^3
- \frac{5559937}{2916} \Nf T_F C_F C_A
- \frac{382033}{648} C_F^2 C_A
\right. \right. \nonumber \\
&& \left. \left. ~~~~
-~ \frac{147473}{81} \zeta_3 C_F C_A^2
- \frac{46219}{27} \zeta_3 C_F^2 C_A
- \frac{28295}{27} \zeta_5 C_F C_A^2
- \frac{12944}{81} \zeta_3 \Nf^2 T_F^2 C_F
\right. \right. \nonumber \\
&& \left. \left. ~~~~
-~ \frac{8000}{9} \zeta_5 \Nf T_F C_F^2
- \frac{7505}{324} \Nf T_F C_F^2
- \frac{640}{9} \zeta_5 \Nf^2 T_F^2 C_F
- \frac{280}{3} \zeta_7 C_F C_A^2
\right. \right. \nonumber \\
&& \left. \left. ~~~~
-~ \frac{44}{3} \zeta_4 \Nf T_F C_F C_A
- \frac{31}{6} C_F^3
- \frac{22}{3} \zeta_4 C_F^2 C_A
+ \frac{22}{3} \zeta_4 C_F C_A^2
+ \frac{44}{3} \zeta_4 \Nf T_F C_F^2
\right. \right. \nonumber \\
&& \left. \left. ~~~~
+~ \frac{64}{3} \zeta_3^2 \Nf T_F C_F C_A
+ \frac{880}{3} \zeta_3^2 C_F C_A^2
+ \frac{16480}{27} \zeta_5 \Nf T_F C_F C_A
+ \frac{18610}{9} \zeta_5 C_F^2 C_A
\right. \right. \nonumber \\
&& \left. \left. ~~~~
+~ \frac{24848}{27} \zeta_3 \Nf T_F C_F^2
+ \frac{83150}{81} \zeta_3 \Nf T_F C_F C_A
+ \frac{196513}{729} \Nf^2 T_F^2 C_F
- 1120 \zeta_7 C_F^3
\right. \right. \nonumber \\
&& \left. \left. ~~~~
+~ \frac{34499767}{11664} C_F C_A^2
- 128 \zeta_3^2 \Nf T_F C_F^2
\right] a^3 \right] \Nc ~+~ O(a^4)
\end{eqnarray}
with the correlation function of the total derivative operator having 
similar properties since we determined that
\begin{eqnarray}
\tilde{\Pi}_{22\,(1)}(a) &=& 
\tilde{\Pi}_{22\,(2)}(a) ~=~ O(a^4) \nonumber \\
\tilde{\Pi}_{22\,(3)}(a) &=& 
2 \tilde{\Pi}_{12\,(3)}(a) ~+~ O(a^4)
\end{eqnarray}
are satisfied explicitly. We note again that the previous three loop 
expressions of \cite{34} were obtained. For $SU(3)$ the numerical values are
\begin{eqnarray}
\left. \tilde{\Pi}_{11\,(1)}(a) \right|^{SU(3)} &=& 1.440000 - 16.734709 a 
+ [ 35.100784 \Nf - 266.402517 ] a^2 \nonumber \\
&& +~ [ - 62.788489 \Nf^2 + 2032.623629 \Nf - 8854.35522 ] a^3 ~+~ O(a^4) 
\nonumber \\
\left. \tilde{\Pi}_{11\,(2)}(a) \right|^{SU(3)} &=& 1.226667 - 10.791041 a 
+ [ 24.025430 \Nf - 173.185423 ] a^2 \nonumber \\
&& +~ [ - 42.292785 \Nf^2 + 1358.239289 \Nf - 5803.326335 ] a^3 ~+~ O(a^4) 
\nonumber \\
\left. \tilde{\Pi}_{11\,(3)}(a) \right|^{SU(3)} &=& 0.226667 + 12.715132 a 
+ [ - 32.978302 \Nf + 225.081859 ] a^2 \nonumber \\
&& +~ [ 48.044306 \Nf^2 - 1598.947786 \Nf + 6915.909940 ] a^3 ~+~ O(a^4) 
\nonumber \\
\left. \tilde{\Pi}_{12\,(3)}(a) \right|^{SU(3)} &=& 3.333333 - 1.799154 a 
+ [ - 10.177094 \Nf + 57.801168 ] a^2 \nonumber \\
&& +~ [ 3.736412 \Nf^2 - 169.803559 \Nf + 1191.052470 ] a^3 ~+~ O(a^4) 
\end{eqnarray}
for the four independent form factors.

\sect{Discussion.}

The main achievement of the article is the determination of the operator matrix
elements for the flavour non-singlet twist-$2$ Wilson operators with moment 
$n$~$=$~$2$ that are relevant for structure functions to four loops. This 
extends the earlier activity of \cite{23} to a new order. Knowledge of these 
matrix elements is necessary to assist with parallel lattice field theory 
evaluations of the same quantity which ultimately have to match the continuum 
high energy behaviour determined here. The extra order of precision in the 
chiral limit at large energy will prove important for reducing uncertainties in
the extraction of forward parton distribution functions. Central to finding the
four loop expressions was the use of the {\sc Forcer} package, \cite{12,13}, 
which contains efficient algorithms to evaluate the underlying Feynman graphs 
in dimensional regularization. While we have provided the analytic expression 
for the main operator matrix element as a function of the linear covariant 
gauge parameter and arbitrary colour group, we have considered related momentum
flow configurations in order to be complete. As the operator matrix element 
depends on the gauge parameter, lattice measurements of it would need to be 
carried out using a gauge fixed action. To address the potential contamination 
of the properties of the operator that are necessary for hadron structure we 
have provided the four loop result of the correlation function of the set of 
$n$~$=$~$2$ moment Wilson operators. Such Green's functions are independent of 
the gauge parameter since the two operators, which mix under renormalization, 
are gauge independent. The analytic expressions will assist lattice matching 
where gauge fixing can be avoided. Throughout we have concentrated on producing
expressions in the $\MSbar$ scheme. In doing so we have provided operator 
matrix elements in a complete format in that the full Lorentz structure is 
available. We have not introduced or discussed the determination of these 
matrix elements in schemes that are more popularly used on the lattice such as 
the $\RI$ scheme of \cite{29,30}. This is because the definition of the $\RI$ 
evaluation of the matrix element is not unambiguous. It depends on the 
combination of Lorentz basis tensors that are chosen to project out that part 
of the matrix element used to define the scheme in the first place. There are 
numerous ways of achieving this. Therefore providing the full Lorentz 
decomposition of the $\MSbar$ operator matrix element is sufficient and 
flexible for any $\RI$ lattice field theory evaluation to be matched to the 
$\MSbar$ result since the latter scheme is the default one for expressing 
physical parameters in. Finally in completing this four loop operator 
renormalization central to lattice matching for unpolarized flavour non-singlet
operators the natural avenue for the four loop programme to turn to next will 
be operators relating to hadronic spin properties such as transversity.

\vspace{1cm}
\noindent
{\bf Data Availability Statement.} The data representing the main results here
are accessible in electronic form from the arXiv ancillary directory associated
with the article.

\vspace{0.3cm}
\noindent
{\bf Competing Interests.} The author has no competing interests to declare 
that are relevant to the content of this article.

\vspace{0.3cm}
\noindent
{\bf Acknowledgements.} The author thanks R. Horsley and P.E.L. Rakow for
constructive discussions and supportive comments. This work was carried out 
with the support of the STFC Consolidated Grant ST/X000699/1 and was undertaken
on Barkla, part of the High Performance Computing facilities at the University 
of Liverpool, UK. For the purpose of open access, the author has applied a
Creative Commons Attribution (CC-BY) licence to any Author Accepted Manuscript 
version arising. 

\appendix

\sect{Form factors for asymmetric momentum routing.}

In this appendix we record analytic expressions for the remaining independent
Landau gauge form factors of the two Green's function where the external 
momentum is channelled through the inserted operator and one quark leg. First 
we have
\begin{eqnarray}
\left. \widehat{\Sigma}^{(1)}_1(p) \right|_{p^2 = \mu^2}^{\alpha = 0} &=&
1 - \frac{32}{9} C_F a
+ \left[
 \frac{1721}{162} \Nf T_F C_F
+ \frac{9259}{648} C_F^2
+ 5 \zeta_3 C_F C_A
- \frac{9527}{324} C_F C_A
\right] a^2
\nonumber \\
&&
+ \left[
22 \zeta_4 C_F^2 C_A
- \frac{26963657}{69984} C_F C_A^2
- \frac{138527}{2916} C_F^3
- \frac{68746}{2187} \Nf^2 T_F^2 C_F
\right. \nonumber \\
&& \left. ~~~~
-~ \frac{40387}{4374} \Nf T_F C_F^2
- \frac{9316}{81} \zeta_3 C_F^3
- \frac{1456}{27} \zeta_3 \Nf T_F C_F^2
- \frac{463}{48} \zeta_4 C_F C_A^2
\right. \nonumber \\
&& \left. ~~~~
-~ \frac{340}{3} \zeta_5 C_F^2 C_A
- \frac{128}{81} \zeta_3 \Nf^2 T_F^2 C_F
- \frac{32}{3} \zeta_4 C_F^3
- \frac{32}{3} \zeta_4 \Nf T_F C_F^2
\right. \nonumber \\
&& \left. ~~~~
+~ \frac{32}{3} \zeta_4 \Nf T_F C_F C_A
+ \frac{305}{12} \zeta_5 C_F C_A^2
+ \frac{320}{3} \zeta_5 C_F^3
+ \frac{452}{81} \zeta_3 \Nf T_F C_F C_A
\right. \nonumber \\
&& \left. ~~~~
+~ \frac{3194}{27} \zeta_3 C_F^2 C_A
+ \frac{40199}{648} \zeta_3 C_F C_A^2
+ \frac{1039685}{4374} \Nf T_F C_F C_A
\right. \nonumber \\
&& \left. ~~~~
+~ \frac{4105721}{17496} C_F^2 C_A
\right] a^3
\nonumber \\
&&
+ \left[
\frac{5951148217}{1259712} C_F^2 C_A^2
- \frac{6055784671}{839808} C_F C_A^3
- \frac{330032317}{314928} \Nf T_F C_F^2 C_A
\right. \nonumber \\
&& \left. ~~~~
-~ \frac{199715627}{279936} C_F^4
- \frac{134713747}{314928} \Nf T_F C_F^3
- \frac{37486813}{157464} \Nf^2 T_F^2 C_F^2
\right. \nonumber \\
&& \left. ~~~~
-~ \frac{15832687}{8748} \Nf^2 T_F^2 C_F C_A
- \frac{4726711}{1458} \zeta_3 \Nf T_F C_F^2 C_A
- \frac{1620881}{13824} \zeta_4 C_F C_A^3
\right. \nonumber \\
&& \left. ~~~~
-~ \frac{1109431}{216} \zeta_3 C_F^3 C_A
- \frac{529823}{96} \zeta_7 C_F^2 C_A^2
- \frac{237905}{864} \zeta_3^2 C_F C_A^3
\right. \nonumber \\
&& \left. ~~~~
-~ \frac{206657}{162} \zeta_5 \Nf T_F C_F C_A^2
- \frac{165845}{864} \zeta_5 \frac{d_F^{abcd} d_A^{abcd}}{\Nc}
- \frac{140875}{576} \zeta_6 \frac{d_F^{abcd} d_A^{abcd}}{\Nc}
\right. \nonumber \\
&& \left. ~~~~
-~ \frac{83144}{27} \zeta_5 C_F^3 C_A
- \frac{80999}{288} \zeta_3^2 \frac{d_F^{abcd} d_A^{abcd}}{\Nc}
- \frac{29153}{27} \zeta_5 C_F^2 C_A^2
\right. \nonumber \\
&& \left. ~~~~
-~ \frac{23975}{32} \zeta_7 \frac{d_F^{abcd} d_A^{abcd}}{\Nc}
- \frac{19292}{81} \zeta_3 \Nf^2 T_F^2 C_F C_A
- \frac{11869}{108} \frac{d_F^{abcd} d_A^{abcd}}{\Nc}
\right. \nonumber \\
&& \left. ~~~~
-~ \frac{9244}{27} \zeta_5 \Nf T_F C_F^3
- \frac{6320}{27} \zeta_4 \Nf T_F C_F^2 C_A
- \frac{4580}{27} \zeta_3 \Nf \frac{d_F^{abcd} d_F^{abcd}}{\Nc}
\right. \nonumber \\
&& \left. ~~~~
-~ \frac{4150}{27} \zeta_6 \Nf T_F C_F C_A^2
- \frac{3100}{3} \zeta_6 C_F^3 C_A
- \frac{2860}{27} \zeta_5 \Nf \frac{d_F^{abcd} d_F^{abcd}}{\Nc}
\right. \nonumber \\
&& \left. ~~~~
-~ \frac{2404}{3} \zeta_3^2 C_F^3 C_A
- \frac{784}{27} \zeta_3^2 \Nf T_F C_F C_A^2
- \frac{362}{3} \zeta_4 C_F^4
- \frac{208}{9} \zeta_3^2 \Nf \frac{d_F^{abcd} d_F^{abcd}}{\Nc}
\right. \nonumber \\
&& \left. ~~~~
-~ \frac{64}{3} \zeta_4 \Nf \frac{d_F^{abcd} d_F^{abcd}}{\Nc}
- \frac{64}{27} \zeta_4 \Nf^3 T_F^3 C_F
- \frac{56}{9} \zeta_3^2 \Nf T_F C_F^2 C_A
+ \frac{64}{3} \zeta_5 \Nf^2 T_F^2 C_F^2
\right. \nonumber \\
&& \left. ~~~~
+~ \frac{80}{243} \zeta_3 \Nf^3 T_F^3 C_F
+ \frac{160}{3} \zeta_3^2 \Nf T_F C_F^3
+ \frac{380}{9} \zeta_4 \Nf T_F C_F^3
+ \frac{400}{3} \zeta_6 \Nf T_F C_F^3
\right. \nonumber \\
&& \left. ~~~~
+~ \frac{700}{9} \zeta_6 \Nf T_F C_F^2 C_A
+ \frac{800}{9} \zeta_6 \Nf \frac{d_F^{abcd} d_F^{abcd}}{\Nc}
+ \frac{1472}{3} \zeta_3^2 C_F^4
+ \frac{1600}{3} \zeta_6 C_F^4
\right. \nonumber \\
&& \left. ~~~~
+~ \frac{2081}{192} \zeta_4 \frac{d_F^{abcd} d_A^{abcd}}{\Nc}
+ \frac{2793}{8} \zeta_7 \Nf T_F C_F C_A^2
+ \frac{3862}{27} \zeta_4 C_F^3 C_A
+ \frac{4366}{9} \zeta_3^2 C_F^2 C_A^2
\right. \nonumber \\
&& \left. ~~~~
+~ \frac{8053}{48} \zeta_4 \Nf T_F C_F C_A^2
+ \frac{10408}{27} \Nf \frac{d_F^{abcd} d_F^{abcd}}{\Nc}
+ \frac{11530}{3} \zeta_5 C_F^4
+ \frac{24125}{72} \zeta_6 C_F^2 C_A^2
\right. \nonumber \\
&& \left. ~~~~
+~ \frac{25900}{3} \zeta_7 C_F^3 C_A
+ \frac{40138}{27} \zeta_5 \Nf T_F C_F^2 C_A
+ \frac{61531}{432} \zeta_4 C_F^2 C_A^2
\right. \nonumber \\
&& \left. ~~~~
+~ \frac{86089}{81} \zeta_3 \Nf T_F C_F^3
+ \frac{381283}{432} \zeta_3 \frac{d_F^{abcd} d_A^{abcd}}{\Nc}
+ \frac{405356}{729} \zeta_3 \Nf^2 T_F^2 C_F^2
\right. \nonumber \\
&& \left. ~~~~
+~ \frac{413225}{6912} \zeta_6 C_F C_A^3
+ \frac{509503}{1296} \zeta_3 \Nf T_F C_F C_A^2
+ \frac{1033771}{6561} \Nf^3 T_F^3 C_F
\right. \nonumber \\
&& \left. ~~~~
+~ \frac{1075115}{1458} \zeta_3 C_F^4
+ \frac{1234681}{1536} \zeta_7 C_F C_A^3
+ \frac{1852697}{1296} \zeta_5 C_F C_A^3
+ \frac{8629757}{15552} \zeta_3 C_F C_A^3
\right. \nonumber \\
&& \left. ~~~~
+~ \frac{19405091}{2916} \Nf T_F C_F C_A^2
+ \frac{30079351}{5832} \zeta_3 C_F^2 C_A^2
+ \frac{128891329}{629856} C_F^3 C_A
\right. \nonumber \\
&& \left. ~~~~
-~ 4704 \zeta_7 C_F^4
- 294 \zeta_7 \Nf T_F C_F^2 C_A
- 20 \zeta_4 \Nf^2 T_F^2 C_F C_A
+ 20 \zeta_4 \Nf^2 T_F^2 C_F^2
\right. \nonumber \\
&& \left. ~~~~
+~ 72 \zeta_5 \Nf^2 T_F^2 C_F C_A
\right] a^4 ~+~ O(a^5) \nonumber \\
\left. \widehat{\Sigma}^{(2)}_1(p) \right|_{p^2 = \mu^2}^{\alpha = 0} &=&
\frac{8}{3} C_F a
+ \left[
\frac{80}{3} C_F C_A
- \frac{98}{9} C_F^2
- \frac{80}{9} \Nf T_F C_F
- 4 \zeta_3 C_F C_A
\right] a^2
\nonumber \\
&&
+ \left[
\frac{1821383}{3888} C_F C_A^2
- \frac{70877}{243} \Nf T_F C_F C_A
- \frac{11257}{54} C_F^2 C_A
- \frac{1388}{9} \zeta_3 C_F^2 C_A
\right. \nonumber \\
&& \left. ~~~~
-~ \frac{911}{9} \zeta_3 C_F C_A^2
- \frac{640}{3} \zeta_5 C_F^3
- \frac{125}{3} \zeta_5 C_F C_A^2
+ \frac{112}{9} \zeta_3 \Nf T_F C_F C_A
\right. \nonumber \\
&& \left. ~~~~
+~ \frac{320}{9} \zeta_3 \Nf T_F C_F^2
+ \frac{640}{3} \zeta_5 C_F^2 C_A
+ \frac{884}{81} \Nf T_F C_F^2
+ \frac{1784}{9} \zeta_3 C_F^3
\right. \nonumber \\
&& \left. ~~~~
+~ \frac{9235}{243} C_F^3
+ \frac{9920}{243} \Nf^2 T_F^2 C_F
\right] a^3
\nonumber \\
&&
+ \left[
\frac{168947053}{17496} C_F C_A^3
- \frac{1097296067}{209952} C_F^2 C_A^2
- \frac{50965121}{5832} \Nf T_F C_F C_A^2
\right. \nonumber \\
&& \left. ~~~~
-~ \frac{13095155}{972} \zeta_3 C_F^2 C_A^2
- \frac{6298873}{2304} \zeta_7 C_F C_A^3
- \frac{1611136}{243} \zeta_3 C_F^4
\right. \nonumber \\
&& \left. ~~~~
-~ \frac{1125245}{1728} \zeta_5 C_F C_A^3
- \frac{465280}{2187} \Nf^3 T_F^3 C_F
- \frac{175888}{81} \zeta_3 \Nf T_F C_F^3
\right. \nonumber \\
&& \left. ~~~~
-~ \frac{120512}{243} \zeta_3 \Nf^2 T_F^2 C_F^2
- \frac{83594}{3} \zeta_7 C_F^3 C_A
- \frac{69760}{9} \zeta_5 C_F^4
- \frac{42127}{18} \zeta_3 \frac{d_F^{abcd} d_A^{abcd}}{\Nc}
\right. \nonumber \\
&& \left. ~~~~
-~ \frac{40899}{32} \zeta_3 C_F C_A^3
- \frac{15160}{9} \zeta_5 \Nf \frac{d_F^{abcd} d_F^{abcd}}{\Nc}
- \frac{6167}{6} \zeta_3^2 C_F^2 C_A^2
\right. \nonumber \\
&& \left. ~~~~
-~ \frac{4263}{4} \zeta_7 \Nf T_F C_F C_A^2
- \frac{1480}{27} \zeta_3 \Nf^2 T_F^2 C_F C_A
- \frac{824}{9} \Nf \frac{d_F^{abcd} d_F^{abcd}}{\Nc}
\right. \nonumber \\
&& \left. ~~~~
-~ \frac{400}{3} \zeta_3^2 \Nf T_F C_F^2 C_A
- \frac{335}{18} \zeta_4 C_F^2 C_A^2
- \frac{128}{9} \zeta_4 C_F^4
- \frac{128}{9} \zeta_4 \Nf T_F C_F^3
\right. \nonumber \\
&& \left. ~~~~
-~ \frac{5}{18} \frac{d_F^{abcd} d_A^{abcd}}{\Nc}
+ \frac{112}{3} \zeta_4 C_F^3 C_A
+ \frac{128}{9} \zeta_4 \Nf T_F C_F^2 C_A
+ \frac{188}{3} \zeta_3^2 \Nf T_F C_F C_A^2
\right. \nonumber \\
&& \left. ~~~~
+~ \frac{2240}{27} \zeta_5 \Nf^2 T_F^2 C_F C_A
+ \frac{5536}{3} \zeta_3^2 C_F^3 C_A
+ \frac{7195}{12} \zeta_5 \Nf T_F C_F C_A^2
\right. \nonumber \\
&& \left. ~~~~
+~ \frac{12995}{32} \zeta_3^2 C_F C_A^3
+ \frac{19360}{9} \zeta_5 \Nf T_F C_F^3
+ \frac{20320}{9} \zeta_3 \Nf \frac{d_F^{abcd} d_F^{abcd}}{\Nc}
\right. \nonumber \\
&& \left. ~~~~
+~ \frac{28721}{48} \zeta_7 \frac{d_F^{abcd} d_A^{abcd}}{\Nc}
+ \frac{63460}{27} \zeta_5 C_F^3 C_A
+ \frac{87325}{72} \zeta_5 \frac{d_F^{abcd} d_A^{abcd}}{\Nc}
\right. \nonumber \\
&& \left. ~~~~
+~ \frac{124600}{27} \zeta_5 C_F^2 C_A^2
+ \frac{381307}{216} \zeta_3 \Nf T_F C_F C_A^2
+ \frac{607718}{243} \zeta_3 \Nf T_F C_F^2 C_A
\right. \nonumber \\
&& \left. ~~~~
+~ \frac{707441}{48} \zeta_7 C_F^2 C_A^2
+ \frac{785233}{6561} \Nf T_F C_F^3
+ \frac{1054445}{54} \zeta_3 C_F^3 C_A
+ \frac{1317287}{2916} C_F^4
\right. \nonumber \\
&& \left. ~~~~
+~ \frac{1784180}{729} \Nf^2 T_F^2 C_F C_A
+ \frac{2098016}{6561} \Nf^2 T_F^2 C_F^2
+ \frac{5323972}{6561} \Nf T_F C_F^2 C_A
\right. \nonumber \\
&& \left. ~~~~
+~ \frac{9835355}{13122} C_F^3 C_A
- 3640 \zeta_5 \Nf T_F C_F^2 C_A
- 992 \zeta_3^2 C_F^4
- 672 \zeta_3^2 \Nf \frac{d_F^{abcd} d_F^{abcd}}{\Nc}
\right. \nonumber \\
&& \left. ~~~~
+~ 631 \zeta_3^2 \frac{d_F^{abcd} d_A^{abcd}}{\Nc}
+ 2352 \zeta_7 \Nf T_F C_F^2 C_A
+ 16464 \zeta_7 C_F^4
\right] a^4 ~+~ O(a^5)
\end{eqnarray}
for the main operator and
\begin{eqnarray} 
\left. \widehat{\Sigma}^{(1)}_2(p) \right|_{p^2 = \mu^2}^{\alpha = 0}
&=& 1
- 2 C_F a
+ \left[
\frac{41}{18} \Nf T_F C_F
+ \frac{67}{8} C_F^2
- \frac{323}{36} C_F C_A
- \zeta_3 C_F C_A
\right] a^2 \nonumber \\
&&
+ \left[
\frac{124}{9} \zeta_3 \Nf T_F C_F C_A
+ \frac{518}{3} \zeta_3 C_F^2 C_A
- \frac{59315}{648} C_F C_A^2
- \frac{5473}{72} \zeta_3 C_F C_A^2
\right. \nonumber \\
&& \left. ~~~~
-~ \frac{523}{6} C_F^3
- \frac{496}{3} \zeta_3 C_F^3
- \frac{430}{81} \Nf^2 T_F^2 C_F
- \frac{323}{18} \Nf T_F C_F^2
- \frac{69}{16} \zeta_4 C_F C_A^2
\right. \nonumber \\
&& \left. ~~~~
-~ \frac{16}{3} \zeta_3 \Nf T_F C_F^2
+ \frac{640}{3} \zeta_5 C_F^3
+ \frac{1075}{12} \zeta_5 C_F C_A^2
+ \frac{3769}{81} \Nf T_F C_F C_A
\right. \nonumber \\
&& \left. ~~~~
+~ \frac{10765}{72} C_F^2 C_A
- 220 \zeta_5 C_F^2 C_A
+ 6 \zeta_4 C_F^2 C_A
\right] a^3 \nonumber \\
&&
+ \left[
\frac{1617}{2} \zeta_7 \Nf T_F C_F C_A^2
- \frac{159880663}{93312} C_F C_A^3
- \frac{976405}{432} \Nf T_F C_F^2 C_A
\right. \nonumber \\
&& \left. ~~~~
-~ \frac{908483}{72} \zeta_3 C_F^3 C_A
- \frac{333785}{972} \Nf^2 T_F^2 C_F C_A
- \frac{311663}{96} \zeta_3 C_F C_A^3
\right. \nonumber \\
&& \left. ~~~~
-~ \frac{305887}{216} \zeta_5 \Nf T_F C_F C_A^2
- \frac{122675}{96} \zeta_5 \frac{d_F^{abcd} d_A^{abcd}}{\Nc}
- \frac{73013}{32} C_F^3 C_A
\right. \nonumber \\
&& \left. ~~~~
-~ \frac{37933}{3} \zeta_7 C_F^2 C_A^2
- \frac{29000}{9} \zeta_5 C_F^3 C_A
- \frac{21860}{9} \zeta_5 \Nf T_F C_F^3
\right. \nonumber \\
&& \left. ~~~~
-~ \frac{20725}{256} \zeta_6 C_F C_A^3
- \frac{18627}{512} \zeta_4 C_F C_A^3
- \frac{12025}{64} \zeta_3^2 C_F C_A^3
\right. \nonumber \\
&& \left. ~~~~
-~ \frac{9905}{16} \zeta_7 \frac{d_F^{abcd} d_A^{abcd}}{\Nc}
- \frac{6979}{2} \zeta_5 C_F^2 C_A^2
- \frac{6847}{32} \zeta_3^2 \frac{d_F^{abcd} d_A^{abcd}}{\Nc}
\right. \nonumber \\
&& \left. ~~~~
-~ \frac{4275}{64} \zeta_6 \frac{d_F^{abcd} d_A^{abcd}}{\Nc}
- \frac{3668}{3} \zeta_3 \Nf \frac{d_F^{abcd} d_F^{abcd}}{\Nc}
- \frac{2437}{2} \zeta_3 \Nf T_F C_F^2 C_A
\right. \nonumber \\
&& \left. ~~~~
-~ \frac{1765}{6} \frac{d_F^{abcd} d_A^{abcd}}{\Nc}
- \frac{1480}{27} \zeta_5 \Nf^2 T_F^2 C_F C_A
- \frac{1125}{16} \zeta_4 C_F^2 C_A^2
\right. \nonumber \\
&& \left. ~~~~
-~ \frac{88}{3} \zeta_3 \Nf^2 T_F^2 C_F C_A
- \frac{16}{9} \zeta_3 \Nf^3 T_F^3 C_F
- \frac{9}{16} \zeta_4 \Nf T_F C_F C_A^2
\right. \nonumber \\
&& \left. ~~~~
+~ \frac{1678}{3} \Nf \frac{d_F^{abcd} d_F^{abcd}}{\Nc}
+ \frac{1720}{3} \zeta_5 \Nf \frac{d_F^{abcd} d_F^{abcd}}{\Nc}
+ \frac{3339}{64} \zeta_4 \frac{d_F^{abcd} d_A^{abcd}}{\Nc}
\right. \nonumber \\
&& \left. ~~~~
+~ \frac{3925}{8} \zeta_6 C_F^2 C_A^2
+ \frac{4413}{4} \zeta_3^2 C_F^2 C_A^2
+ \frac{11134}{3} \zeta_5 \Nf T_F C_F^2 C_A
\right. \nonumber \\
&& \left. ~~~~
+~ \frac{15137}{9} \zeta_3 \Nf T_F C_F^3
+ \frac{20845}{6} \zeta_3 C_F^4
+ \frac{22387}{729} \Nf^3 T_F^3 C_F
\right. \nonumber \\
&& \left. ~~~~
+~ \frac{29589}{16} \zeta_3 \frac{d_F^{abcd} d_A^{abcd}}{\Nc}
+ \frac{38125}{384} C_F^4
+ \frac{41521}{72} \zeta_3 \Nf T_F C_F C_A^2
\right. \nonumber \\
&& \left. ~~~~
+~ \frac{62551}{6} \zeta_3 C_F^2 C_A^2
+ \frac{67256}{3} \zeta_7 C_F^3 C_A
+ \frac{106655}{144} \Nf T_F C_F^3
\right. \nonumber \\
&& \left. ~~~~
+~ \frac{110065}{648} \Nf^2 T_F^2 C_F^2
+ \frac{367703}{192} \zeta_7 C_F C_A^3
+ \frac{5734315}{3888} \Nf T_F C_F C_A^2
\right. \nonumber \\
&& \left. ~~~~
+~ \frac{8843773}{3456} \zeta_5 C_F C_A^3
+ \frac{10913573}{2592} C_F^2 C_A^2
- 12936 \zeta_7 C_F^4
- 1876 \zeta_3^2 C_F^3 C_A
\right. \nonumber \\
&& \left. ~~~~
-~ 1470 \zeta_7 \Nf T_F C_F^2 C_A
- 900 \zeta_6 C_F^3 C_A
- 150 \zeta_4 C_F^4
- 112 \zeta_3^2 \Nf T_F C_F^2 C_A
\right. \nonumber \\
&& \left. ~~~~
-~ 50 \zeta_6 \Nf T_F C_F C_A^2
- 24 \zeta_4 \Nf T_F C_F^3
- 12 \zeta_4 \Nf^2 T_F^2 C_F^2
\right. \nonumber \\
&& \left. ~~~~
+~ 12 \zeta_4 \Nf^2 T_F^2 C_F C_A
+ 12 \zeta_3 \Nf^2 T_F^2 C_F^2
+ 50 \zeta_3^2 \Nf T_F C_F C_A^2
\right. \nonumber \\
&& \left. ~~~~
+~ 100 \zeta_6 \Nf T_F C_F^2 C_A
+ 256 \zeta_3^2 \Nf \frac{d_F^{abcd} d_F^{abcd}}{\Nc}
+ 306 \zeta_4 C_F^3 C_A
\right. \nonumber \\
&& \left. ~~~~
+~ 400 \zeta_6 C_F^4
+ 1040 \zeta_3^2 C_F^4
+ 7310 \zeta_5 C_F^4
\right] a^4 ~+~ O(a^5) \nonumber \\
\left. \widehat{\Sigma}^{(2)}_2(p) \right|_{p^2 = \mu^2}^{\alpha = 0}
&=& 
4 C_F a
+ \left[
\frac{346}{9} C_F C_A
- \frac{104}{9} \Nf T_F C_F
- 18 C_F^2
- 4 \zeta_3 C_F C_A
\right] a^2 \nonumber \\
&&
+ \left[
\frac{86}{9} \Nf T_F C_F^2
- \frac{31312}{81} \Nf T_F C_F C_A
- \frac{3439}{9} C_F^2 C_A
- \frac{1280}{3} \zeta_5 C_F^3
\right. \nonumber \\
&& \left. ~~~~
-~ \frac{986}{9} \zeta_3 C_F C_A^2
- \frac{772}{3} \zeta_3 C_F^2 C_A
- \frac{290}{3} \zeta_5 C_F C_A^2
+ \frac{64}{9} \zeta_3 \Nf T_F C_F C_A
\right. \nonumber \\
&& \left. ~~~~
+~ \frac{128}{3} \zeta_3 \Nf T_F C_F^2
+ \frac{973}{6} C_F^3
+ \frac{992}{3} \zeta_3 C_F^3
+ \frac{4000}{81} \Nf^2 T_F^2 C_F
\right. \nonumber \\
&& \left. ~~~~
+~ \frac{54643}{81} C_F C_A^2
+ 400 \zeta_5 C_F^2 C_A
\right] a^3 \nonumber \\
&&
+ \left[
\frac{40743547}{2916} C_F C_A^3
- \frac{2839646}{243} \Nf T_F C_F C_A^2
- \frac{921776}{81} C_F^2 C_A^2
\right. \nonumber \\
&& \left. ~~~~
-~ \frac{333291}{64} \zeta_7 C_F C_A^3
- \frac{173120}{729} \Nf^3 T_F^3 C_F
- \frac{165067}{3} \zeta_7 C_F^3 C_A
\right. \nonumber \\
&& \left. ~~~~
-~ \frac{64865}{54} \zeta_5 C_F C_A^3
- \frac{32512}{3} \zeta_3 C_F^4
- \frac{32128}{9} \zeta_3 \Nf T_F C_F^3
\right. \nonumber \\
&& \left. ~~~~
-~ \frac{21560}{3} \zeta_5 \Nf T_F C_F^2 C_A
- \frac{17329}{18} \zeta_3 C_F C_A^3
- \frac{10880}{3} \zeta_5 \Nf \frac{d_F^{abcd} d_F^{abcd}}{\Nc}
\right. \nonumber \\
&& \left. ~~~~
-~ \frac{10610}{9} \Nf T_F C_F^3
- \frac{10435}{12} C_F^4
- \frac{1664}{3} \zeta_3 \Nf^2 T_F^2 C_F^2
\right. \nonumber \\
&& \left. ~~~~
-~ \frac{512}{3} \Nf \frac{d_F^{abcd} d_F^{abcd}}{\Nc}
- \frac{69}{4} \zeta_4 C_F^2 C_A^2
+ \frac{400}{3} \frac{d_F^{abcd} d_A^{abcd}}{\Nc}
\right. \nonumber \\
&& \left. ~~~~
+~ \frac{511}{8} \zeta_7 \frac{d_F^{abcd} d_A^{abcd}}{\Nc}
+ \frac{5600}{27} \zeta_5 \Nf^2 T_F^2 C_F C_A
+ \frac{5705}{8} \zeta_3^2 C_F C_A^3
\right. \nonumber \\
&& \left. ~~~~
+~ \frac{8090}{3} \zeta_3 \Nf T_F C_F^2 C_A
+ \frac{9995}{3} \zeta_5 \frac{d_F^{abcd} d_A^{abcd}}{\Nc}
+ \frac{14080}{3} \zeta_3 \Nf \frac{d_F^{abcd} d_F^{abcd}}{\Nc}
\right. \nonumber \\
&& \left. ~~~~
+~ \frac{24602}{9} \zeta_3 \Nf T_F C_F C_A^2
+ \frac{28645}{27} \zeta_5 \Nf T_F C_F C_A^2
+ \frac{29585}{4} \zeta_5 C_F^2 C_A^2
\right. \nonumber \\
&& \left. ~~~~
+~ \frac{32132}{81} \Nf^2 T_F^2 C_F^2
+ \frac{43840}{9} \zeta_5 \Nf T_F C_F^3
+ \frac{50650}{9} \zeta_5 C_F^3 C_A
\right. \nonumber \\
&& \left. ~~~~
+~ \frac{66151}{27} \Nf T_F C_F^2 C_A
+ \frac{234023}{36} C_F^3 C_A
+ \frac{599857}{18} \zeta_3 C_F^3 C_A
\right. \nonumber \\
&& \left. ~~~~
+~ \frac{690613}{24} \zeta_7 C_F^2 C_A^2
+ \frac{728416}{243} \Nf^2 T_F^2 C_F C_A
- 22359 \zeta_3 C_F^2 C_A^2
\right. \nonumber \\
&& \left. ~~~~
-~ 16560 \zeta_5 C_F^4
- 3622 \zeta_3 \frac{d_F^{abcd} d_A^{abcd}}{\Nc}
- 2382 \zeta_3^2 C_F^2 C_A^2
+ 736 \zeta_3^2 \frac{d_F^{abcd} d_A^{abcd}}{\Nc}
\right. \nonumber \\
&& \left. ~~~~
-~ 2058 \zeta_7 \Nf T_F C_F C_A^2
- 1280 \zeta_3^2 \Nf \frac{d_F^{abcd} d_F^{abcd}}{\Nc}
- 128 \zeta_3 \Nf^2 T_F^2 C_F C_A
\right. \nonumber \\
&& \left. ~~~~
-~ 96 \zeta_3^2 \Nf T_F C_F^2 C_A
+ 24 \zeta_4 C_F^3 C_A
+ 68 \zeta_3^2 \Nf T_F C_F C_A^2
+ 4064 \zeta_3^2 C_F^3 C_A
\right. \nonumber \\
&& \left. ~~~~
-~ 2240 \zeta_3^2 C_F^4
+ 4704 \zeta_7 \Nf T_F C_F^2 C_A
+ 32928 \zeta_7 C_F^4
\right] a^4
\,+\, O(a^5)
\end{eqnarray}
for the total derivative operator. Finally the remaining form factors for the
other configuration in the Landau gauge are 
\begin{eqnarray}
\left. \tilde{\Sigma}^{(1)}_1(p) \right|_{p^2 = \mu^2}^{\alpha = 0} &=&
-~ \frac{14}{9} C_F a
+ \left[
6 \zeta_3 C_F C_A
- \frac{1655}{81} C_F C_A
+ \frac{479}{81} C_F^2
+ \frac{676}{81} \Nf T_F C_F
\right] a^2
\nonumber \\
&&
+~ \left[
16 \zeta_4 C_F^2 C_A
- \frac{20557637}{69984} C_F C_A^2
- \frac{57136}{2187} \Nf^2 T_F^2 C_F
- \frac{1468}{27} \zeta_3 C_F^2 C_A
\right. \nonumber \\
&& \left. ~~~~
-~ \frac{1312}{27} \zeta_3 \Nf T_F C_F^2
- \frac{664}{81} \zeta_3 \Nf T_F C_F C_A
- \frac{385}{6} \zeta_5 C_F C_A^2
- \frac{320}{3} \zeta_5 C_F^3
\right. \nonumber \\
&& \left. ~~~~
-~ \frac{128}{81} \zeta_3 \Nf^2 T_F^2 C_F
- \frac{32}{3} \zeta_4 C_F^3
- \frac{32}{3} \zeta_4 \Nf T_F C_F^2
- \frac{16}{3} \zeta_4 C_F C_A^2
\right. \nonumber \\
&& \left. ~~~~
+~ \frac{32}{3} \zeta_4 \Nf T_F C_F C_A
+ \frac{320}{3} \zeta_5 C_F^2 C_A
+ \frac{4076}{81} \zeta_3 C_F^3
+ \frac{11182}{81} \zeta_3 C_F C_A^2
\right. \nonumber \\
&& \left. ~~~~
+~ \frac{19051}{2187} \Nf T_F C_F^2
+ \frac{115651}{2916} C_F^3
+ \frac{744913}{8748} C_F^2 C_A
+ \frac{836159}{4374} \Nf T_F C_F C_A
\right] a^3
\nonumber \\
&&
+~ \left[
8232 \zeta_7 C_F^4
- \frac{288553669}{52488} C_F C_A^3
- \frac{45996029}{39366} \Nf T_F C_F^3
- \frac{30720221}{5832} \zeta_3 C_F^2 C_A^2
\right. \nonumber \\
&& \left. ~~~~
-~ \frac{14219297}{17496} C_F^4
- \frac{11709743}{10368} \zeta_5 C_F C_A^3
- \frac{8029076}{19683} \Nf^2 T_F^2 C_F^2
\right. \nonumber \\
&& \left. ~~~~
-~ \frac{6414311}{4374} \Nf^2 T_F^2 C_F C_A
- \frac{1995110}{729} \zeta_3 C_F^4
- \frac{1475069}{729} \zeta_3 \Nf T_F C_F^2 C_A
\right. \nonumber \\
&& \left. ~~~~
-~ \frac{568981}{512} \zeta_7 C_F C_A^3
- \frac{237875}{1296} \zeta_3 \Nf T_F C_F C_A^2
- \frac{151135}{1728} \zeta_3^2 C_F C_A^3
\right. \nonumber \\
&& \left. ~~~~
-~ \frac{104405}{108} \zeta_3 \frac{d_F^{abcd} d_A^{abcd}}{\Nc}
- \frac{60068}{27} \zeta_5 \Nf T_F C_F^2 C_A
- \frac{50144}{81} \zeta_3 \Nf T_F C_F^3
\right. \nonumber \\
&& \left. ~~~~
-~ \frac{41356}{3} \zeta_7 C_F^3 C_A
- \frac{22253}{36} \zeta_3^2 C_F^2 C_A^2
- \frac{18340}{27} \zeta_5 \Nf \frac{d_F^{abcd} d_F^{abcd}}{\Nc}
\right. \nonumber \\
&& \left. ~~~~
-~ \frac{16916}{81} \zeta_3 \Nf^2 T_F^2 C_F C_A
- \frac{10400}{3} \zeta_5 C_F^4
- \frac{6320}{27} \zeta_4 \Nf T_F C_F^2 C_A
\right. \nonumber \\
&& \left. ~~~~
-~ \frac{4694}{27} \Nf \frac{d_F^{abcd} d_F^{abcd}}{\Nc}
- \frac{4400}{27} \zeta_4 C_F^3 C_A
- \frac{4367}{54} \zeta_4 C_F C_A^3
- \frac{4165}{32} \zeta_7 \frac{d_F^{abcd} d_A^{abcd}}{\Nc}
\right. \nonumber \\
&& \left. ~~~~
-~ \frac{3675}{8} \zeta_7 \Nf T_F C_F C_A^2
- \frac{2800}{27} \zeta_6 \Nf T_F C_F C_A^2
- \frac{2512}{9} \zeta_3^2 \Nf \frac{d_F^{abcd} d_F^{abcd}}{\Nc}
\right. \nonumber \\
&& \left. ~~~~
-~ \frac{2134}{27} \zeta_3^2 \Nf T_F C_F C_A^2
- \frac{1648}{3} \zeta_3^2 C_F^4
- \frac{1600}{9} \zeta_6 \frac{d_F^{abcd} d_A^{abcd}}{\Nc}
- \frac{1400}{9} \zeta_6 C_F^2 C_A^2
\right. \nonumber \\
&& \left. ~~~~
-~ \frac{1211}{18} \zeta_3^2 \frac{d_F^{abcd} d_A^{abcd}}{\Nc}
- \frac{400}{3} \zeta_6 C_F^3 C_A
- \frac{200}{9} \zeta_6 \Nf T_F C_F^2 C_A
- \frac{124}{3} \zeta_4 \frac{d_F^{abcd} d_A^{abcd}}{\Nc}
\right. \nonumber \\
&& \left. ~~~~
-~ \frac{64}{3} \zeta_4 \Nf \frac{d_F^{abcd} d_F^{abcd}}{\Nc}
- \frac{64}{27} \zeta_4 \Nf^3 T_F^3 C_F
+ \frac{64}{3} \zeta_5 \Nf^2 T_F^2 C_F^2
+ \frac{88}{3} \zeta_4 C_F^4
\right. \nonumber \\
&& \left. ~~~~
+~ \frac{160}{3} \zeta_3^2 \Nf T_F C_F^3
+ \frac{400}{3} \zeta_6 C_F^4
+ \frac{400}{3} \zeta_6 \Nf T_F C_F^3
+ \frac{505}{3} \zeta_4 \Nf T_F C_F C_A^2
\right. \nonumber \\
&& \left. ~~~~
+~ \frac{512}{243} \zeta_3 \Nf^3 T_F^3 C_F
+ \frac{596}{9} \zeta_4 \Nf T_F C_F^3
+ \frac{800}{9} \zeta_6 \Nf \frac{d_F^{abcd} d_F^{abcd}}{\Nc}
\right. \nonumber \\
&& \left. ~~~~
+~ \frac{952}{9} \zeta_3^2 \Nf T_F C_F^2 C_A
+ \frac{3224}{3} \zeta_3^2 C_F^3 C_A
+ \frac{3424}{27} \zeta_5 \Nf^2 T_F^2 C_F C_A
+ \frac{3800}{27} \zeta_6 C_F C_A^3
\right. \nonumber \\
&& \left. ~~~~
+~ \frac{3856}{27} \zeta_5 C_F^3 C_A
+ \frac{19901}{108} \frac{d_F^{abcd} d_A^{abcd}}{\Nc}
+ \frac{28432}{27} \zeta_3 \Nf \frac{d_F^{abcd} d_F^{abcd}}{\Nc}
\right. \nonumber \\
&& \left. ~~~~
+~ \frac{45953}{216} \zeta_4 C_F^2 C_A^2
+ \frac{56336}{27} \zeta_5 \Nf T_F C_F^3
+ \frac{91033}{648} \zeta_5 \Nf T_F C_F C_A^2
\right. \nonumber \\
&& \left. ~~~~
+~ \frac{130127}{54} \zeta_5 C_F^2 C_A^2
+ \frac{228011}{32} \zeta_7 C_F^2 C_A^2
+ \frac{396608}{729} \zeta_3 \Nf^2 T_F^2 C_F^2
\right. \nonumber \\
&& \left. ~~~~
+~ \frac{469115}{432} \zeta_5 \frac{d_F^{abcd} d_A^{abcd}}{\Nc}
+ \frac{808009}{108} \zeta_3 C_F^3 C_A
+ \frac{832288}{6561} \Nf^3 T_F^3 C_F
\right. \nonumber \\
&& \left. ~~~~
+~ \frac{23860433}{19683} \Nf T_F C_F^2 C_A
+ \frac{48937694}{19683} C_F^3 C_A
+ \frac{59119163}{15552} \zeta_3 C_F C_A^3
\right. \nonumber \\
&& \left. ~~~~
+~ \frac{60417419}{11664} \Nf T_F C_F C_A^2
+ \frac{647151739}{1259712} C_F^2 C_A^2
- 32 \zeta_4 \Nf^2 T_F^2 C_F C_A
\right. \nonumber \\
&& \left. ~~~~
+~ 32 \zeta_4 \Nf^2 T_F^2 C_F^2
+ 1176 \zeta_7 \Nf T_F C_F^2 C_A
\right] a^4 ~+~ O(a^5) \nonumber \\
\left. \tilde{\Sigma}^{(2)}_1(p) \right|_{p^2 = \mu^2}^{\alpha = 0} &=&
-~ \frac{4}{3} C_F a
+ \left[
\frac{64}{9} C_F^2
- \frac{106}{9} C_F C_A
+ \frac{8}{3} \Nf T_F C_F
\right] a^2
\nonumber \\
&&
+~ \left[
55 \zeta_5 C_F C_A^2
- \frac{801481}{3888} C_F C_A^2
- \frac{60343}{486} C_F^3
- \frac{2080}{243} \Nf^2 T_F^2 C_F
- \frac{1192}{9} \zeta_3 C_F^3
\right. \nonumber \\
&& \left. ~~~~
-~ \frac{560}{3} \zeta_5 C_F^2 C_A
- \frac{64}{9} \zeta_3 \Nf T_F C_F^2
+ \frac{16}{3} \zeta_3 \Nf T_F C_F C_A
+ \frac{25}{3} \zeta_3 C_F C_A^2
\right. \nonumber \\
&& \left. ~~~~
+~ \frac{110}{81} \Nf T_F C_F^2
+ \frac{640}{3} \zeta_5 C_F^3
+ \frac{928}{9} \zeta_3 C_F^2 C_A
+ \frac{9377}{54} C_F^2 C_A
\right. \nonumber \\
&& \left. ~~~~
+~ \frac{23059}{243} \Nf T_F C_F C_A
\right] a^3
\nonumber \\
&&
+~ \left[
1248 \zeta_3^2 C_F^4
- \frac{150932057}{26244} C_F^3 C_A
- \frac{75514229}{17496} C_F C_A^3
- \frac{10750721}{6561} \Nf T_F C_F^2 C_A
\right. \nonumber \\
&& \left. ~~~~
- \frac{504676}{6561} \Nf^2 T_F^2 C_F^2
- \frac{401068}{729} \Nf^2 T_F^2 C_F C_A
- \frac{372563}{27} \zeta_3 C_F^3 C_A
\right. \nonumber \\
&& \left. ~~~~
-~ \frac{300395}{108} \zeta_5 C_F^2 C_A^2
- \frac{224595}{16} \zeta_7 C_F^2 C_A^2
- \frac{209141}{216} \zeta_3 \Nf T_F C_F C_A^2
\right. \nonumber \\
&& \left. ~~~~
-~ \frac{152555}{72} \zeta_5 \frac{d_F^{abcd} d_A^{abcd}}{\Nc}
- \frac{90827}{288} \zeta_3 C_F C_A^3
- \frac{88490}{27} \zeta_5 C_F^3 C_A
\right. \nonumber \\
&& \left. ~~~~
-~ \frac{49825}{108} \zeta_5 \Nf T_F C_F C_A^2
- \frac{47572}{243} \zeta_3 \Nf T_F C_F^2 C_A
- \frac{21920}{9} \zeta_3 \Nf \frac{d_F^{abcd} d_F^{abcd}}{\Nc}
\right. \nonumber \\
&& \left. ~~~~
-~ \frac{9825}{32} \zeta_3^2 C_F C_A^3
- \frac{6656}{3} \zeta_3^2 C_F^3 C_A
- \frac{2405}{18} \frac{d_F^{abcd} d_A^{abcd}}{\Nc}
- \frac{1120}{9} \zeta_5 \Nf^2 T_F^2 C_F C_A
\right. \nonumber \\
&& \left. ~~~~
-~ \frac{128}{9} \zeta_4 C_F^4
- \frac{128}{9} \zeta_4 \Nf T_F C_F^3
- \frac{112}{3} \zeta_3^2 \Nf T_F C_F^2 C_A
- \frac{49}{36} \zeta_4 C_F^2 C_A^2
\right. \nonumber \\
&& \left. ~~~~
-~ \frac{16}{3} \zeta_3^2 \Nf T_F C_F C_A^2
+ \frac{40}{3} \zeta_4 C_F^3 C_A
+ \frac{128}{9} \zeta_4 \Nf T_F C_F^2 C_A
+ \frac{712}{9} \Nf \frac{d_F^{abcd} d_F^{abcd}}{\Nc}
\right. \nonumber \\
&& \left. ~~~~
+~ \frac{1976}{27} \zeta_3 \Nf^2 T_F^2 C_F C_A
+ \frac{3969}{4} \zeta_7 \Nf T_F C_F C_A^2
+ \frac{8125}{6} \zeta_3^2 C_F^2 C_A^2
\right. \nonumber \\
&& \left. ~~~~
+~ \frac{10640}{3} \zeta_5 \Nf T_F C_F^2 C_A
+ \frac{14272}{243} \zeta_3 \Nf^2 T_F^2 C_F^2
+ \frac{17480}{9} \zeta_5 \Nf \frac{d_F^{abcd} d_F^{abcd}}{\Nc}
\right. \nonumber \\
&& \left. ~~~~
+~ \frac{23069}{18} \zeta_3 \frac{d_F^{abcd} d_A^{abcd}}{\Nc}
+ \frac{25655}{48} \zeta_7 \frac{d_F^{abcd} d_A^{abcd}}{\Nc}
+ \frac{54080}{2187} \Nf^3 T_F^3 C_F
\right. \nonumber \\
&& \left. ~~~~
+~ \frac{79280}{9} \zeta_5 C_F^4
+ \frac{81473}{3} \zeta_7 C_F^3 C_A
+ \frac{113264}{81} \zeta_3 \Nf T_F C_F^3
+ \frac{950435}{1728} \zeta_5 C_F C_A^3
\right. \nonumber \\
&& \left. ~~~~
+~ \frac{963248}{729} C_F^4
+ \frac{1022336}{243} \zeta_3 C_F^4
+ \frac{5699603}{2304} \zeta_7 C_F C_A^3
+ \frac{8519923}{6561} \Nf T_F C_F^3
\right. \nonumber \\
&& \left. ~~~~
+~ \frac{8637793}{972} \zeta_3 C_F^2 C_A^2
+ \frac{17186383}{5832} \Nf T_F C_F C_A^2
+ \frac{1291947325}{209952} C_F^2 C_A^2
\right. \nonumber \\
&& \left. ~~~~
-~ 16464 \zeta_7 C_F^4
- 2720 \zeta_5 \Nf T_F C_F^3
- 2352 \zeta_7 \Nf T_F C_F^2 C_A
- 105 \zeta_3^2 \frac{d_F^{abcd} d_A^{abcd}}{\Nc}
\right. \nonumber \\
&& \left. ~~~~
+~ 608 \zeta_3^2 \Nf \frac{d_F^{abcd} d_F^{abcd}}{\Nc}
\right] a^4 ~+~ O(a^5) ~.
\end{eqnarray}

\end{document}